\documentclass[prd,twocolumn,a4paper,floatfix,nofootinbib,preprintnumbers,superscriptaddress]{revtex4-1}
\usepackage[utf8]{inputenc}
\usepackage[table]{xcolor}
\usepackage{float}
\usepackage{amsthm}
\usepackage{graphicx}
\usepackage{amsfonts}
\usepackage{booktabs}
\usepackage{siunitx}
\usepackage{mathtools}
\usepackage{multirow}
\usepackage{array}
\usepackage[normalem]{ulem}
\usepackage{orcidlink}
\usepackage{natbib}
\usepackage{bm}
\usepackage{cleveref}
\usepackage{caption}
\usepackage{tabularx}
\usepackage{subcaption}
\usepackage[justification=RaggedRight]{caption}
\usepackage{transparent}
\usepackage[multiple]{footmisc}
\usepackage{import}

\usepackage[nohyperlinks, nolist]{acronym}
\usepackage{hyperref}

\graphicspath{{./}}

\newcommand{\boldtheta}{\boldsymbol{\theta}}

\newcommand{\MTOV}{M_{\rm{TOV}}}
\newcommand{\src}{{\rm{src}}}
\newcommand{\comp}{{\rm{comp}}}
\newcommand{\LambdaTilde}{\tilde{\Lambda}}
\newcommand{\DeltaLambdaTilde}{\delta\tilde{\Lambda}}
\newcommand{\Msun}{{\rm{M}}_{\odot}}

\newcommand{\chiEFT}{{{\chi_{\rm{EFT}}}}}

\definecolor{jeffreysred1}{HTML}{f6cdb0}
\definecolor{jeffreysred2}{HTML}{fc9074}
\definecolor{jeffreysred3}{HTML}{f4744c}
\definecolor{jeffreysred4}{HTML}{ef5b43}
\definecolor{jeffreysred5}{HTML}{f52a44}

\begin{document}

\title{Incorporating neutron star physics into gravitational wave inference with physics-informed priors using normalizing flows}

\author{{Thibeau Wouters\thanks{t.r.i.wouters@uu.nl}~\orcidlink{0009-0006-2797-3808}}}
\email{t.r.i.wouters@uu.nl}
\affiliation{Institute for Gravitational and Subatomic Physics (GRASP), Utrecht University, Princetonplein 1, 3584 CC Utrecht, The Netherlands}
\affiliation{Nikhef, Science Park 105, 1098 XG Amsterdam, The Netherlands}
\author{{Peter T. H. Pang~\orcidlink{0000-0001-7041-3239}}}
\affiliation{Nikhef, Science Park 105, 1098 XG Amsterdam, The Netherlands}
\affiliation{Institute for Gravitational and Subatomic Physics (GRASP), Utrecht University, Princetonplein 1, 3584 CC Utrecht, The Netherlands}
\author{{Tim Dietrich~\orcidlink{0000-0003-2374-307X}}}
\affiliation{Institut f\"ur Physik und Astronomie, Universit\"at Potsdam, Haus 28, Karl-Liebknecht-Str. 24/25, 14476, Potsdam, Germany}
\affiliation{Max Planck Institute for Gravitational Physics (Albert Einstein Institute), Am M\"{u}hlenberg 1, Potsdam 14476, Germany}
\author{{Chris Van Den Broeck~\orcidlink{0000-0001-6800-4006}}}
\affiliation{Institute for Gravitational and Subatomic Physics (GRASP), Utrecht University, Princetonplein 1, 3584 CC Utrecht, The Netherlands}
\affiliation{Nikhef, Science Park 105, 1098 XG Amsterdam, The Netherlands}
\date{\today}

\begin{abstract}
Bayesian inference, widely used in gravitational-wave parameter estimation, depends on the choice of priors, i.e., on our previously existing knowledge.
However, to investigate neutron star mergers, priors are often chosen in an agnostic way, leaving valuable information from nuclear physics and independent observations of neutron stars unused.
In this work, we propose to encode information on neutron star physics into physics-informed prior distributions constructed with normalizing flows.
These priors take input from constraints on the nuclear equation of state and neutron star mass distributions.
Applied to GW170817, GW190425, and GW230529, we highlight two contributions of the framework.
First, we demonstrate its ability to provide source classification and to enable model selection of equation of state constraints for loud signals such as GW170817, directly from the gravitational-wave data.
Second, we obtain narrower constraints on the source properties through these informed priors.
As a result, these physics-informed priors consistently recover higher luminosity distances compared to agnostic priors.
Our method provides a scalable way for classifying future ambiguous low-mass mergers observed through gravitational waves and for informing single-event gravitational-wave data analysis with neutron star physics.
\end{abstract}

\maketitle

\acrodef{AD}[AD]{automatic differentiation}
\acrodef{JIT}[JIT]{just-in-time}
\acrodef{JSD}[JSD]{Jensen-Shannon divergence}
\acrodef{KL}[KL]{Kullback-Leibler divergence}
\acrodef{PE}[PE]{parameter estimation}
\acrodef{ET}[ET]{Einstein Telescope}
\acrodef{CE}[CE]{Cosmic Explorer}
\acrodef{MCMC}[MCMC]{Markov chain Monte Carlo}
\acrodef{GW}[GW]{gravitational wave}
\acrodef{LVK}[LVK]{LIGO-Virgo-KAGRA}
\acrodef{EM}[EM]{electromagnetic}
\acrodef{CBC}[CBC]{compact binary coalescence}
\acrodef{PN}[PN]{post-Newtonian}
\acrodef{NS}[NS]{neutron star}
\acrodef{PSR}[PSR]{pulsar}
\acrodef{KDE}[KDE]{kernel density estimate}
\acrodef{NF}[NF]{normalizing flow}
\acrodef{BBH}[BBH]{binary black hole}
\acrodef{BNS}[BNS]{binary neutron star}
\acrodef{NSBH}[NSBH]{neutron star-black hole}
\acrodef{EOS}[EOS]{equation of state}
\acrodef{EFT}[EFT]{effective field theory}
\acrodef{NEP}[NEP]{nuclear empirical parameter}
\acrodef{HIC}[HIC]{heavy-ion collision}
\acrodef{MM}[MM]{metamodel}
\acrodef{CSE}[CSE]{speed-of-sound extension scheme}
\acrodef{TOV}[TOV]{Tolman-Oppenheimer-Volkoff}
\acrodef{JS}[JS]{Jensen-Shannon}
\acrodef{CPU}[CPU]{central processing unit}
\acrodef{GPU}[GPU]{graphical processing unit}
\acrodef{TPU}[TPU]{tensor processing unit}
\acrodef{ML}[ML]{machine learning}
\acrodef{SNR}[SNR]{signal-to-noise ratio}
\acrodef{PSD}[PSD]{power spectral density}
\acrodef{NICER}[NICER]{Neutron star Interior Composition ExploreR}
\acrodef{3G}[3G]{third-generation}
\acrodef{BH}[BH]{black hole}

\acresetall

\section{Introduction}\label{sec:introduction}

Observing \acp{CBC} through \acp{GW} offers a unique way to probe the properties of compact objects in the Universe~\cite{LIGOScientific:2018mvr, LIGOScientific:2020ibl, LIGOScientific:2021usb, KAGRA:2021vkt, LIGOScientific:2025slb}.
After the first part of the fourth observing run of the \ac{LVK} Collaborations~\cite{KAGRA:2013rdx, LIGO:2024kkz, Capote:2024rmo, LIGOScientific:2025slb}, more than $200$ confident detections of \acp{CBC} involving \acp{BH} and \acp{NS} have been reported by Advanced LIGO~\cite{LIGOScientific:2014pky}, Advanced Virgo~\cite{VIRGO:2014yos}, and KAGRA~\cite{KAGRA:2020tym}.
The characterization of these sources relies on Bayesian inference~\cite{Veitch:2008ur, Veitch:2009hd, Veitch:2012df, Veitch:2014wba}, i.e., on the interplay between the constraining power of the observed data and on our existing prior knowledge.
In current analyses, it is common to adopt agnostic priors that do not encode any astrophysical assumption~\cite{LIGOScientific:2025yae}.
However, agnostic priors neglect the growing body of information on compact objects from both theory and observations outside the field of \ac{GW} astronomy.

In the case of \acp{NS}, for instance, the macroscopic properties are dictated by the supranuclear \ac{EOS}~\cite{Lattimer:2000nx, Lattimer:2012nd, Ozel:2016oaf, Burgio:2021vgk, Chatziioannou:2024tjq}.
This has two important implications for \ac{GW} data analysis.
First, \acp{NS} are tidally deformed in a binary due to the gravitational field of their companion, which modifies the \ac{GW} signal with respect to the point particle baseline~\cite{Flanagan:2007ix, Damour:2009vw, Hinderer:2009ca, Damour:2012yf}.
The magnitude of this deformation for a given \ac{NS} mass depends on the \ac{EOS}, such that constraints on the \ac{EOS} provide non-trivial prior knowledge on the tidal deformabilities of \acp{NS}.
Second, the \ac{EOS} predicts a maximum mass for non-rotating, spherically symmetric \acp{NS}, the \ac{TOV} mass ($\MTOV$)~\cite{Tolman:1939jz, Oppenheimer:1939ne}.
Since this informs us about the transition from \ac{NS} to \ac{BH} masses, it can be used in source classification, for instance, to distinguish between \ac{BNS} and \ac{NSBH} mergers.
Beyond these \ac{EOS}-dependent properties, the mass distribution of \acp{NS} can be informed by observations of galactic binaries using radio timing~\cite{Demorest:2010bx, Hobbs:2006cd} and pulse profile modeling~\cite{Bogdanov:2019ixe}, providing information complementary to the \ac{EOS} constraints.

Today, a diverse set of \ac{EOS} constraints already exists from nuclear theory and experiments, and observations of \acp{NS} (see Ref.~\cite{Koehn:2024set} for an overview).
However, this body of knowledge is expected to grow substantially with more advanced \ac{EM} observatories~\cite{eXTP:2018anb, Watts:2018iom, STROBE-XScienceWorkingGroup:2019cyd} and upcoming ground-based \ac{GW} detectors~\cite{Punturo:2010zza, ET:2019dnz, ETScienceCase2020, ETDesignUpdate2024}.
Therefore, the community requires a framework to coherently incorporate this information of \ac{NS} properties into \ac{GW} inference.

One solution is to perform joint or hierarchical inference to constrain hyperparameters of a population model with a collection of observations~\cite{Thrane:2018qnx}, as commonly done for population studies~\cite{LIGOScientific:2018jsj,LIGOScientific:2020kqk,KAGRA:2021duu,LIGOScientific:2025pvj}.
Past works have already performed joint, hierarchical inference on mass distributions and the \ac{EOS}~\cite{Wysocki:2020myz,Golomb:2021tll,Golomb:2024lds,Biswas:2024hja,Ghosh:2024cwc}.
However, due to the rapid increase in dimensionality of the parameter space, this is prohibitively expensive to be applied directly on the \ac{GW} strain data, and instead takes single-event \ac{GW} posterior samples as input~\cite{LIGOScientific:2025pvj} (although see a recent advancement for \ac{BBH} populations by Ref.~\cite{Leyde:2026hvm}).

An alternative, complementary solution is to incorporate current knowledge on \ac{NS} properties into the prior distributions for \ac{GW} data analysis, using posterior predictive distributions obtained from a hierarchical inference.
Since population hyperparameters are not updated during sampling, this method keeps the dimensionality fixed, allowing it to be used directly on the strain data.
As a consequence, we can obtain informed posterior distributions that are conditioned on our knowledge of \ac{NS} properties prior to analyzing a new \ac{GW} event.
For instance,
Ref.~\cite{De:2018uhw} uses mass distributions inspired by \ac{NS} populations.
However, while they enforce \acp{NS} to follow a common \ac{EOS}, this is done through a universal relation without incorporating existing \ac{EOS} constraints.
Ref.~\cite{Magnall:2025zhm} explicitly models the correlation between the source-frame chirp mass and the mass-weighted tidal deformability~\cite{Flanagan:2007ix, Favata:2013rwa, Wade:2014vqa} based on \ac{EOS} constraints.
However, they do not employ informed priors for the second tidal deformability parameter for \ac{BNS} systems or for the masses.
Therefore, past work has so far considered only a subset of the relevant \ac{NS} physics into priors for \ac{GW} data analysis.

\begin{figure}[t]
    \centering
    \includegraphics[width=\columnwidth]{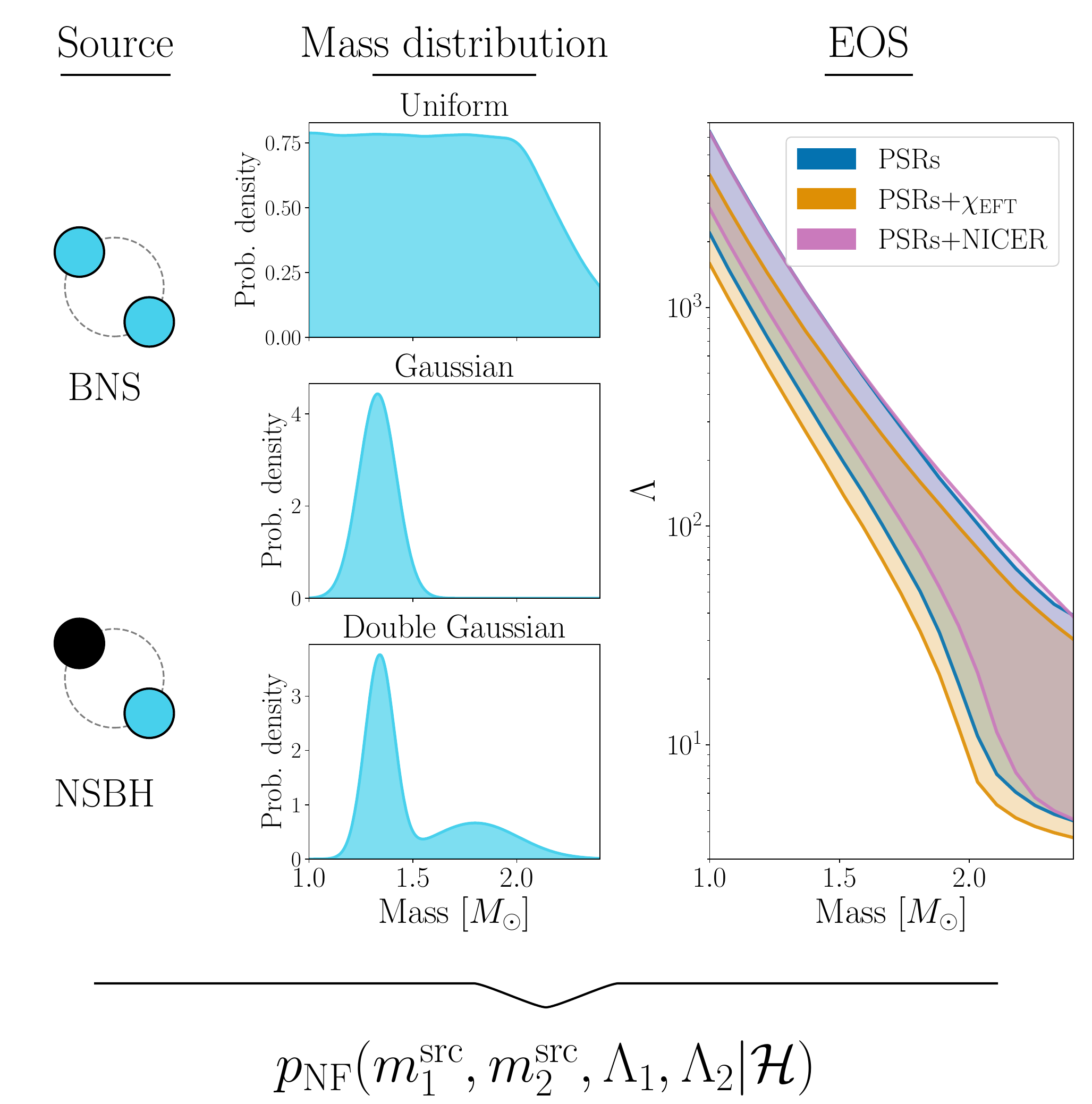}
    \caption{Schematic overview of the physics-informed priors.
    For both \ac{BNS} and \ac{NSBH} binaries (left column), a training dataset of masses and tidal deformabilities is generated, where the former are obtained by assuming a specific \ac{NS} mass distribution (middle panel) and the latter from a set of \ac{EOS} constraints (right panel).
    A normalizing flow (NF) is trained on these samples to render their distribution tractable.
    The learned densities are then used as priors in Bayesian inference.
    Further details are given in Sec.~\ref{sec: methods: neural priors construction details}.
    }
    \label{fig: Figure 1}
\end{figure}

In this work, we introduce a flexible framework to simultaneously incorporate information from both \ac{NS} mass distributions and \ac{EOS} constraints into \ac{GW} data analysis.
The approach, illustrated schematically in Fig.~\ref{fig: Figure 1}, generates training data by drawing source-frame masses from mass distributions and predicting tidal deformabilities from sets of viable \acp{EOS}.
We train \acp{NF}, a type of neural density estimators, on these samples to create tractable, physics-informed prior distributions.
These priors enable Bayesian model selection between competing hypotheses for a single event based on source type, mass distribution, and \ac{EOS} constraints.
Such model selection routines are important to understand ambiguous \ac{GW} events, where components fall in the `mass gap' between \ac{NS} and \ac{BH} masses, such as GW230529~\cite{LIGOScientific:2024elc}.
By embedding these constraints in a joint prior on masses and tidal deformabilities, we can obtain physics-informed posteriors directly from the \ac{GW} strain data.
The data-driven nature of our framework allows it to readily be updated with future advances in our understanding of \ac{NS} properties.

This paper is organized as follows.
In Sec.~\ref{sec: methods}, we provide a recap of Bayesian statistics,  introduce the concept of \acp{NF}, and provide details on the input and construction of our physics-informed priors.
By applying the method to GW170817~\cite{LIGOScientific:2017vwq}, GW190425~\cite{LIGOScientific:2020aai}, and GW230529~\cite{LIGOScientific:2024elc}, we show how these physics-informed priors enable us to provide model selection on the nature of the merger in Sec.~\ref{sec: results: model selection}.
At the same time, the informed priors provide narrower posteriors on the source parameters, as shown in Sec.~\ref{sec: results: parameter constraints}.
In Sec.~\ref{sec: discussion}, we provide additional context to our method, before concluding in Sec.~\ref{sec: conclusion}.

\section{Methods}\label{sec: methods}

\subsection{Bayesian inference}\label{sec: methods: Bayesian inference}

Based on Bayes' theorem, the posterior distribution $p(\boldtheta|d, \mathcal{H})$ of parameters $\boldtheta$ conditioned on a hypothesis $\mathcal{H}$ and data $d$ is given by
\begin{equation}\label{eq: Bayes}
    p(\boldtheta | d, \mathcal{H}) = \frac{p(d|\boldtheta, \mathcal{H}) p(\boldtheta|\mathcal{H})}{p(d | \mathcal{H})} \, ,
\end{equation}
where $p(\boldtheta|\mathcal{H})$ is the prior distribution, $p(d|\boldtheta, \mathcal{H})$ is the likelihood function and $p(d| \mathcal{H})$ is the Bayesian evidence.

Since the posterior distribution is intractable in \ac{GW} parameter estimation, it must be obtained through numerical samplers~\cite{Veitch:2008ur, Veitch:2009hd, Veitch:2012df, Veitch:2014wba}.
In this work, we use nested sampling~\cite{Skilling:2004pqw, Skilling:2006gxv, Ashton:2022grj} to obtain the posterior samples and to estimate the Bayesian evidence.
In particular, we use the \textsc{dynesty} sampler~\cite{Speagle:2019ivv, sergey_koposov_2024_12537467} implemented in \textsc{bilby}~\cite{Ashton:2018jfp, Romero-Shaw:2020owr} with $4096$ live points.
For the likelihood evaluation, we use the \texttt{IMRPhenomXP\_NRTidalv3} waveform model~\cite{Colleoni:2023ple, Abac:2023ujg}, and accelerate it with the multibanding technique~\cite{Garcia-Quiros:2020qlt, Morisaki:2021ngj}.

The parameters of the \ac{GW} waveform are detailed in Appendix~\ref{sec: app: GW parameters and priors}, along with the prior choices that are common across the analyses conducted in this work.
We compare the posteriors obtained with our physics-informed priors against those that use the `default' prior used in \ac{GW} parameter estimation.
In particular, this samples the component masses uniformly in a chirp mass range restricted by the \ac{GW} event and samples the component tidal deformabilities uniformly in the range $[0, 5000]$ (see Appendix~\ref{sec: app: GW parameters and priors} for further details).
Throughout the paper, we also refer to these priors as `uninformed' or `agnostic', as they do not assume any prior astrophysical knowledge of \acp{NS}.

The Bayesian evidence, i.e., the normalization constant of the posterior in Eq.~\eqref{eq: Bayes}, provides a quantitative measure of the plausibility between two competing hypotheses $\mathcal{H}_1$ and $\mathcal{H}_2$ through the odds ratio
\begin{equation}
    \mathcal{O}_2^1 = \frac{p(d|\mathcal{H}_1)}{p(d|\mathcal{H}_2)} \frac{p(\mathcal{H}_1)}{p(\mathcal{H}_2)} = \mathcal{B}_2^1 \Pi_2^1 \, ,
\end{equation}
where $\mathcal{B}_2^1$ is the Bayes factor, and $\Pi_2^1$ is the prior odds.
In this work, we assume the prior odds to be unity, such that the Bayes factors are equal to the odds ratio.
Future work could consider using prior odds informed by, for example, the expected rates of \ac{BNS} and \ac{NSBH} mergers~\cite{LIGOScientific:2020kqk, KAGRA:2021duu, LIGOScientific:2025pvj}.
Therefore, the Bayes factor quantifies the relative preference for one hypothesis over another, balancing the goodness-of-fit to the data against the prior volume, which penalizes more flexible models through the Occam's razor principle.

The Bayes factors are reported in $\log_{10}$ scale.
When $\log_{10}(\mathcal{B}_2^1) > 0$, hypothesis $\mathcal{H}_1$ is preferred over $\mathcal{H}_2$.
To interpret the magnitude of the Bayes factors, we will use Jeffreys' scale, shown in Table~\ref{tab:jeffreys_scale}.
The last column introduces a color scheme for the scale that will be used to present our Bayes factor results in Table~\ref{tab: Bayes factors}.

\begin{table}[t]
    \centering
    \caption{Jeffreys' scale for the interpretation of Bayes factors in $\log_{10}$ scale.
    The final column shows the color coding that will be used in Table~\ref{tab: Bayes factors}.
    }
    \label{tab:jeffreys_scale}
    \renewcommand{\arraystretch}{1.2}

    \begin{tabular}{
        >{\centering\arraybackslash}p{0.19\columnwidth}
        >{\centering\arraybackslash}p{0.62\columnwidth}
        >{\centering\arraybackslash}p{0.14\columnwidth}
    }
        \hline\hline
        $\log_{10}(\mathcal{B}_2^1)$ & Interpretation & Color \\
        \hline
        $[0,\, \tfrac{1}{2}]$ & Barely worth mentioning & \cellcolor{jeffreysred1} \\
        $[\tfrac{1}{2},\, 1]$ & Substantial evidence & \cellcolor{jeffreysred2} \\
        $[1,\, \tfrac{3}{2}]$ & Strong evidence & \cellcolor{jeffreysred3} \\
        $[\tfrac{3}{2},\, 2]$ & Very strong evidence & \cellcolor{jeffreysred4} \\
        $> 2$ & Decisive evidence & \cellcolor{jeffreysred5} \\
        \hline\hline
    \end{tabular}
\end{table}

\subsection{Normalizing flows}

Normalizing flows (NFs) are a class of generative machine learning models that enable flexible density estimation, and are therefore often referred to as neural density estimators~\cite{Rezende:2015ocs, Kobyzev:2019ydm, Papamakarios:2019fms}.
The core idea of \acp{NF} is to transform a tractable base distribution $p_{\mathcal{Z}}(z)$ in a latent space (often, a multivariate standard Gaussian) into a complicated target distribution $p_{\rm{\mathcal{X}}}(x)$ in the data space $\mathcal{X}$ through a bijective transformation $f: \mathcal{Z} \to \mathcal{X}$ with a numerically tractable Jacobian.
As such, the density in the data space can be computed with the change of variables formula:
\begin{equation}
    p_{\rm{\mathcal{X}}}(x) = p_{\mathcal{Z}}\left(f^{-1}(x)\right)\left|\det\left( \frac{\partial f^{-1}(x)}{\partial x} \right)\right| \, .
\end{equation}
Besides density evaluation, the bijection can transform samples from the base distribution to samples in the data space.
Being parametrized by neural networks, these transformations are expressive and can be trained with samples from the target distribution.

In \ac{GW} data analysis, \acp{NF} have been widely used to make parameter estimation more efficient, either through likelihood-free inference (see, e.g., Refs~\cite{Green:2020hst, Dax:2021tsq, Ashton:2021anp, Dax:2022pxd, Langendorff:2022fzq, Bhardwaj:2023xph, Sun:2023vlq, Du:2023plr, Dax:2024mcn, Gupte:2024jfe, Vallisneri:2024xfk, Kolmus:2024scm, Xiong:2024gpx, Hu:2024lrj, Lai:2025xov, Lyu:2025vqk, Qin:2025mvj}), by accelerating samplers~\cite{Williams:2021qyt, Williams:2023ppp, Karamanis:2022ksp, Wong:2023lgb, Wouters:2024oxj, Williams:2025aar, Villa:2025ofq}, or through variational inference~\cite{Vallisneri:2024xfk, Mould:2025dts}.
Other works use \acp{NF} to estimate the Bayesian evidence from \ac{GW} posterior samples~\cite{Srinivasan:2024uax, Polanska:2024zpn} or in hierarchical Bayesian inference, for instance, to model population likelihoods~\cite{Wong:2020jdt, Wong:2020yig, Wong:2020ise, Cheung:2021orb, Ruhe:2022ddi, Colloms:2025hib} or the pseudo-likelihoods for inferring the \ac{EOS} from \ac{GW} posteriors of \ac{BNS} mergers~\cite{Wouters:2025zju, Pang:2025fes, Wouters:2025ull}.

Outside of these applications, \acp{NF} are well-suited for constructing data-driven, multi-dimensional joint prior distributions, as they are normalized by design, have tractable densities, and can efficiently generate samples.
This has already been shown in various applications of Bayesian inference in astrophysics~\cite{Alsing:2021wef, Bevins:2022qsc, Alsing:2024tlr, Zhang:2024thl, Mootoovaloo:2024sao, Zhang:2025yco, Thorp:2025bpl, Roch:2025jpu, Hoogkamer:2025eaq}.
In the field of \acp{GW}, Ref.~\cite{Mould:2023ift} used kernel density estimators to approximate priors of \ac{GW} parameters based on simulated sources from population models and recommended \acp{NF} for more complex distributions.
Ref.~\cite{Prathaban:2024rmu} used \acp{NF} to construct a repartitioned prior from a low-resolution nested sampling run.
Ref.~\cite{Villa:2025ygb} obtained an orthogonal reparametrization of signal and noise parameters in Bayesian inference of pulsar timing arrays with \acp{NF}.
Finally, Ref.~\cite{Malz:2025xdg} relied on \acp{NF} to create informed priors on their data-driven glitch model to simulate blip glitches.

In this work, we build on this line of research and use \acp{NF} to parametrize prior distributions over masses and tidal deformabilities for \ac{GW} data analysis informed by \ac{NS} physics.
The next section outlines the input information and the procedure used to construct our priors.

\subsection{Physics-informed priors}\label{sec: methods: neural priors}

\subsubsection{Equation of state constraints}\label{sec: methods: neural priors: EOS}

Our prior expectations for the tidal deformabilities of \acp{NS} are based on some of the existing constraints on the \ac{EOS}~\cite{Koehn:2024set}.
For this, we create sets of \ac{EOS} samples from a hybrid, phenomenological parametrization.
From low to high densities, this parametrization consists of a fixed crust~\cite{Douchin:2001sv}, a metamodel part~\cite{Tews:2018iwm, Margueron:2017eqc, Margueron:2017lup, Somasundaram:2020chb}, and an agnostic expansion scheme based on the speed of sound~\cite{Tews:2018iwm, Greif:2018njt, Tews:2019cap, Somasundaram:2021clp}.
In total, this provides a flexible parametrization, with $26$ parameters being varied on the fly.
Further details are given in Ref.~\cite{Wouters:2025zju}.

We constrain the \ac{EOS} parameters with the following constraints, acting as likelihoods in the Bayesian inference of the \ac{EOS}:
\begin{itemize}
    \item \textbf{PSRs}: Mass measurements of the heaviest \acp{PSR} observed so far, with masses around $\sim 2\,\Msun$, provide a lower bound on the \ac{TOV} mass.
    In practice, we use the measurements of \ac{PSR} J1614-2230~\cite{Demorest:2010bx, Shamohammadi:2022ttx} and \ac{PSR} J0740+6620~\cite{Fonseca:2021wxt}.

    \item \textbf{PSRS+}$\boldsymbol{\chiEFT}$: Chiral effective field theory ($\chiEFT$) expands the nuclear Hamiltonian as a function of the nucleon momenta divided by a breakdown scale and provides predictions for the \ac{EOS} at lower energies~\cite{Epelbaum:2008ga, Machleidt:2011zz}.
    After solving the nuclear many-body calculations, the uncertainties due to the truncation of this expansion can be tracked~\cite{Epelbaum:2014efa, Drischler:2020hwi, Drischler:2020yad, Armstrong:2025tza}.
    As a result, $\chiEFT$ provides an uncertainty estimate that we use to constrain the \ac{EOS} at low densities.
    In particular, we directly take the constraints derived by Ref.~\cite{Koehn:2024set}, and refer readers to this work for further details.

    \item \textbf{PSRs+NICER}: The \ac{NICER} obtains mass-radius measurements of pulsars through pulse profile modeling~\cite{gendreau2016neutron, Bogdanov:2019ixe, riley2023xpsi}.
    Here, we limit ourselves to PSR J0030+0451 and PSR J0740+6620, since their properties have been studied extensively~\cite{Riley:2019yda, Miller:2019cac, Riley:2021pdl, Miller:2021qha, Vinciguerra:2023qxq, Dittmann:2024mbo, Hoogkamer:2025ype}.
    Future work can consider the inclusion of more recent mass-radius measurements~\cite{Choudhury:2024xbk, Salmi:2024bss, Mauviard:2025dmd}.
\end{itemize}
As indicated by the notation, the $\chiEFT$ and \ac{NICER} constraints are each used jointly with the heavy \acp{PSR} constraint.

The posterior \ac{EOS} samples, used in the construction of the physics-informed priors, are obtained by directly sampling our parametrization with the bove constraints.
To achieve this, we use \textsc{jester}~\cite{Wouters:2025zju}, which accelerates the inference with GPUs~\cite{frostig2019compiling} and \ac{NF}-enhanced samplers~\cite{Wong:2022xvh, Gabrie:2021tlu}.
We note that the inclusion of the $\chiEFT$ (\ac{NICER}) constraints results in softer (stiffer) \ac{EOS} posteriors~\cite{Wouters:2025zju}, which will also be reflected in the physics-informed priors (see Fig.~\ref{fig: neural priors}).

\subsubsection{NS mass distributions}\label{sec: methods: neural priors: populations}

There are various models proposed for the mass distribution of \acp{NS} in binary systems.
Measurements of galactic double \ac{NS} systems point to models such as a single Gaussian~\cite{Ozel:2012ax, Kiziltan:2013oja, Ozel:2016oaf} or a double Gaussian~\cite{Schwab:2010jm, Antoniadis:2016hxz, Alsing:2017bbc, Farrow:2019xnc, Shao:2020bzt, Farr_2020, Niu:2025nha}.
However, \ac{NS} masses inferred from \acp{GW} from both \ac{BNS} and \ac{NSBH} mergers are consistent with a uniform distribution~\cite{LIGOScientific:2021qlt, Landry:2021hvl, Golomb:2024lds}.
Therefore, our \ac{NS} mass priors are based on the following three mass distributions:
\begin{itemize}
    \item \textbf{Uniform}: The \ac{NS} masses are drawn from a uniform distribution in the range $[1\,\Msun, \MTOV]$.\footnote{While the maximum mass of an \ac{NS} is increased if it is spinning~\cite{Breu:2016ufb, LIGOScientific:2021qlt, KAGRA:2021duu}, we expect this effect to be negligible for our analyses, since we use low-spin priors for \acp{NS} (see Appendix~\ref{sec: app: GW parameters and priors}).}
    We will detail how the uncertainty in $\MTOV$ is taken into account in Sec.~\ref{sec: methods: neural priors construction details}.
    The lower bound is motivated by the \ac{NS} mass of $\sim 1.17\,\Msun$ observed in Refs.~\cite{Martinez:2015mya, Suwa:2018uni}, but is slightly lowered due to the tentative observation of \acp{NS} with lower mass~\cite{Chen:2025uwd, Doroshenko:2022nwp, Horvath:2023uwl}.

    \item \textbf{Gaussian}: The \ac{NS} masses are drawn from a Gaussian distribution with mean $1.33\,\Msun$ and standard deviation $0.09\,\Msun$~\cite{Ozel:2016oaf}.

    \item \textbf{Double Gaussian}: The \ac{NS} masses are drawn from a weighted mixture of two Gaussian distributions.
    The first has a mean of $1.34\,\Msun$, a standard deviation of $0.07\,\Msun$, and a relative weight\footnote{See Ref.~\cite{Shao:2020bzt} for a definition of the relative weight.} of $0.65$.
    The second has a mean of $1.80\,\Msun$ and standard deviation $0.21\,\Msun$~\cite{Alsing:2017bbc}.
\end{itemize}
For \ac{BNS} systems, we use the above models to generate the source-frame masses $m_1^\src \geq m_2^\src$ with random pairing~\cite{Landry:2021hvl}.

In the case of \ac{NSBH} systems, these models are used to estimate the mass of the secondary component.
For the primary, we restrict our attention to \ac{BH} masses that fall within the lower mass gap~\cite{Bailyn:1997xt, Ozel:2010su, Farr:2010tu}.
However, we choose a minimal mass distribution that only considers information from the \ac{EOS}.
In particular, we sample the \ac{BH} mass uniformly in the range $[\MTOV, 5\,\Msun]$, where the upper bound is chosen conservatively to accommodate the primary mass of GW230529~\cite{LIGOScientific:2024elc}.
We note that the choice of this upper bound determines the lower bound for the mass ratio.
For example, for the uniform mass distribution for \acp{NS}, it is restricted to $q\geq 1/5$, and therefore cannot be used for systems with more unequal masses, such as GW190814~\cite{LIGOScientific:2020zkf}.
However, the events considered in this work have negligible support for primary masses above $5\,\Msun$, such that this choice is expected to have a minimal impact on the posteriors.
Besides, future work can consider \ac{BH} mass distributions that are informed by proposed theoretical models~\cite{Fishbach:2020ryj, Farah:2021qom, Chandra:2024ila, Fishbach:2025bjh, Ray:2025aqr, Baumgarte:2025syh} or observations~\cite{Thompson:2018ycv, LIGOScientific:2020zkf, Jayasinghe:2021uqb, Song:2024tqr, Barr:2024wwl}.

\subsubsection{Construction of physics-informed priors}\label{sec: methods: neural priors construction details}

With the \ac{EOS} constraints from Sec.~\ref{sec: methods: neural priors: EOS} and the mass distributions of Sec.~\ref{sec: methods: neural priors: populations} at hand, we now detail the construction of the physics-informed priors to model the joint distribution of masses and tidal deformabilities.
In total, $18$ different physics-informed priors are constructed by varying the source type (\ac{BNS} or \ac{NSBH}), the \ac{NS} mass distribution (uniform, Gaussian, or double Gaussian), and the set of \ac{EOS} constraints (PSRs, PSRs+$\chiEFT$, or PSRs+NICER).
For each physics-informed prior, the training data is created as follows:
\begin{itemize}
    \item Draw an \ac{EOS} curve, which determines
    the value of $\MTOV$ and the function $\Lambda(m)$ for this sample.

    \item Draw the source-frame masses $m_{1,2}^\src$ according to the specified \ac{NS} mass distribution and source type.
    The uniform \ac{NS} mass model and the \ac{BH} mass distribution (for \ac{NSBH} systems) take the $\MTOV$ value from the current \ac{EOS} sample.

    \item Compute $\Lambda_{1,2}$ according to the \ac{EOS}.
    For the \ac{NSBH} case, we only compute $\Lambda_2 = \Lambda(m_2^\src)$ and enforce $\Lambda_1 = 0$.
    For the \ac{BNS} case, we compute both $\Lambda_1 = \Lambda(m_1^\src)$ and $\Lambda_2 = \Lambda(m_2^\src)$.
\end{itemize}
For \ac{BNS} systems, this produces a $4$-dimensional distribution of samples $p(m_1^\src, m_2^\src, \Lambda_1, \Lambda_2|\mathcal{H})$, where $\mathcal{H}$ encodes the specific hypothesis (\ac{BNS} system, mass distribution, \ac{EOS} constraints).
For \ac{NSBH} systems, on the other hand, this gives a $3$-dimensional distribution $p(m_1^\src, m_2^\src, \Lambda_2|\mathcal{H})$.

\begin{figure*}[t]
    \centering
    \includegraphics[width=\textwidth]{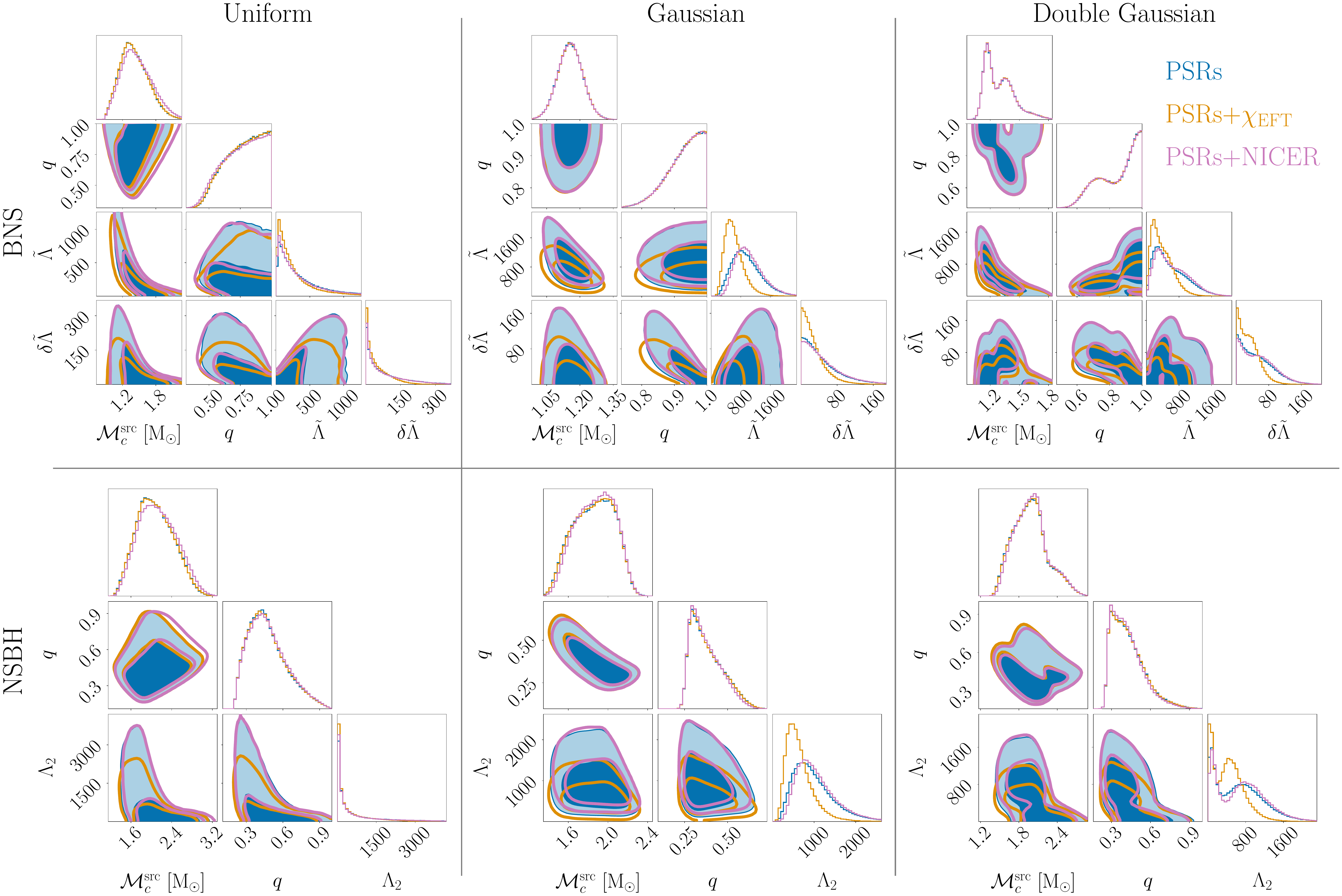}
        \caption{Physics-informed priors on source-frame masses and tidal deformabilities considered in this work.
        The top row (bottom row) shows the priors for \ac{BNS} (\ac{NSBH}) systems, while the columns show different mass distributions and the colors denote the \ac{EOS} constraints.
        The contour lines show the $68\%$ and $95\%$ probability areas of the distributions.
        }
    \label{fig: neural priors}
\end{figure*}

We train \acp{NF} to approximate these distributions and render them tractable to be used as priors in samplers.
Further details on the numerical implementation and training can be found in Appendix~\ref{sec: app: NF}.
The resulting physics-informed priors are shown in Fig.~\ref{fig: neural priors}.
We show the source-frame chirp mass $\mathcal{M}_c^\src$ and mass ratio $q = m_2/m_1$, and (for \ac{BNS} systems) the mass-weighted tidal deformabilities $\LambdaTilde$, $\DeltaLambdaTilde$ (see, e.g., Refs.~\cite{Flanagan:2007ix, Favata:2013rwa, Wade:2014vqa} for a definition), since these are the combinations of parameters that affect the \ac{GW} waveform.

\section{Results}\label{sec: results}

We apply the physics-informed priors in inferences of GW170817, GW190425, and GW230529.
In the following sections, we highlight the ability of the physics-informed priors to perform Bayesian model selection (Sec.~\ref{sec: results: model selection}) and to produce narrow posteriors informed by \ac{NS} physics (Sec.~\ref{sec: results: parameter constraints}).

\subsection{Model selection}\label{sec: results: model selection}

As discussed in Sec.~\ref{sec: methods: Bayesian inference}, we perform model selection by computing Bayes factors between hypotheses encoded as different physics-informed priors.
For each \ac{GW} event, we refer to the hypothesis that achieves the highest Bayesian evidence as the reference model.
This model is marked as `ref.' in the Tab.~\ref{tab: Bayes factors}, and we show the $\log_{10}$-Bayes factors between the reference model and other hypotheses.
The entries are color-coded according to Jeffrey's scale as shown in Tab.~\ref{tab:jeffreys_scale}.
To give further context, Fig.~\ref{fig: combined log likelihood plots} shows the histograms of the log-likelihoods evaluated on the obtained posterior samples for the different physics-informed priors (colors), compared against the posterior obtained from the uninformed prior (gray), where we limit ourselves to results from the source type and mass distribution to which the reference model belongs.

We emphasize that the Bayes factors should be interpreted on the level of the single event being analyzed and not on a population level.
That is, the Bayes factors compare various hypotheses for each event individually and cannot be used to derive constraints regarding the mass distributions or the \ac{EOS} constraints that were used to create the priors that encode these hypotheses.
Put differently, these Bayes factors tell us how well the event can be explained by having the source-frame masses drawn from a certain mass model and the tidal deformabilities inferred from a given set of \ac{EOS} constraints.

\subsubsection{GW170817}

For GW170817, we find overwhelming Bayes factors in favor of the \ac{BNS} hypothesis over the \ac{NSBH} hypothesis, which is in agreement with previous findings, e.g., Refs.~\cite{Hinderer:2018pei, Coughlin:2019kqf}.
This is mainly due to the mass of the signal, since GW170817 has a source-frame chirp mass of around $1.186\,\Msun$, where the \ac{NSBH} physics-informed priors have negligible support (see Fig.~\ref{fig: neural priors}).

Between the different \ac{BNS} hypotheses, the reference model comes from assuming a Gaussian mass distribution.
The double Gaussian and uniform mass distributions are disfavored, with varying degrees of evidence.
Within the Gaussian mass distribution, the physics-informed prior informed by $\chiEFT$ has substantial evidence in favor of it compared to the other physics-informed priors, implying that the GW170817 measurement is consistent with a soft \ac{EOS}.
This observation agrees with inferences on the \ac{EOS} from GW170817~\cite{Annala:2017llu, Radice:2017lry, LIGOScientific:2018cki, Landry:2018prl, Capano:2019eae, Raaijmakers:2019dks, Essick:2019ldf, Radice:2018ozg, Dietrich:2020efo, Huth:2020ozf, Chatziioannou:2020pqz, Pang:2022rzc} and conclusions drawn in other studies~\cite{Lim:2018bkq, Tews:2018iwm, Lim:2019som, Tews:2019cap, Tews:2019ioa, Essick:2020flb, Essick:2021ezp}.
However, our work establishes this conclusion directly from the \ac{GW} data.

As shown in the top panel of Fig.~\ref{fig: combined log likelihood plots}, the higher evidence for $\chiEFT$ arises because this physics-informed prior concentrates more of its density in areas of high likelihood compared to the other priors.
The physics-informed priors trained without $\chi$EFT information cover these regions less well, resulting in overall lower likelihoods.

\subsubsection{GW190425}\label{sec: results: model selection: GW190425}

For GW190425, the Bayes factors in Table~\ref{tab: Bayes factors} show that the \ac{BNS} hypothesis is preferred with substantial to decisive evidence over the \ac{NSBH} hypothesis, depending on the mass distribution and \ac{EOS} constraints.
However, in contrast to GW170817, we cannot definitely rule out the \ac{NSBH} hypothesis across all prior choices, in line with previous works~\cite{Gupta:2019nwj, Han:2020qmn, Foley:2020kus, Kyutoku:2020xka, Takhistov:2020vxs, Barbieri:2020ebt, Dudi:2021abi}.

Assuming a \ac{BNS} origin, the highest evidence is achieved for the uniform mass distribution due to the high masses of GW190425.
This is consistent with the findings of Refs.~\cite{LIGOScientific:2020aai, LIGOScientific:2021qlt, Landry:2021hvl, Golomb:2024lds}, stating that the masses of GW190425 are outliers compared to the population of galactic binaries.
Indeed, the Gaussian, respectively double Gaussian, mass distribution is disfavored with decisive, respectively substantial, evidence.
Contrary to GW170817, there is no evidence discriminating between the different \ac{EOS} constraints.
This can be attributed to the lower \ac{SNR} of GW190425 and the higher masses, which result in lower tidal deformabilities.

The middle panel of Fig.~\ref{fig: combined log likelihood plots} shows that the distributions of likelihood values have a larger median when using any physics-informed prior compared to the uninformed prior.
Therefore, all physics-informed priors lead to a larger number of posterior samples to fall in regions of high likelihood.
Differences among the \ac{EOS} sets are insignificant, such that we can not rule out any models regarding the stiffness of the \ac{EOS}, consistent with the conclusions drawn from the Bayes factors.

\subsubsection{GW230529}\label{sec: results: model selection: GW230529}

The results in Table~\ref{tab: Bayes factors} classify GW230529 as an \ac{NSBH} merger with decisive evidence.
While this conclusion is in agreement with previous works~\cite{LIGOScientific:2024elc, Koehn:2024ape, Janquart:2024ztv, Markin:2025oeo}, our analysis establishes it robustly in a Bayesian context, with consistent support across a range of prior choices.

The Bayes factors between the physics-informed priors are barely worth mentioning.
Besides the low \ac{SNR} of the signal, this can further be attributed to two reasons.
First, we note that all physics-informed priors for the \ac{NSBH} systems shown in Fig.~\ref{fig: neural priors} have similar support for a system with a source-frame chirp mass of around $\sim 2\,\Msun$.
Second, effects from tidal deformabilities are hard to measure in systems like GW230529, since the \texttt{NRTidalv3} waveform tapers off at low frequencies~\cite{Dietrich:2018uni, Abac:2023ujg}.
Indeed, the amplitude of the waveform is tapered at $641_{-158}^{+318}$ Hz when estimated from the uninformed posterior samples, or at $858_{-108}^{+113}$ Hz if computed with posterior samples from the reference model (both at $90\%$ credibility).
Since phase differences due to the tidal deformabilities manifest themselves mainly at higher frequencies, the effects of different \ac{EOS} constraints are hard to distinguish.

\begin{figure}[t]
    \centering
    \includegraphics[width=0.975\columnwidth]{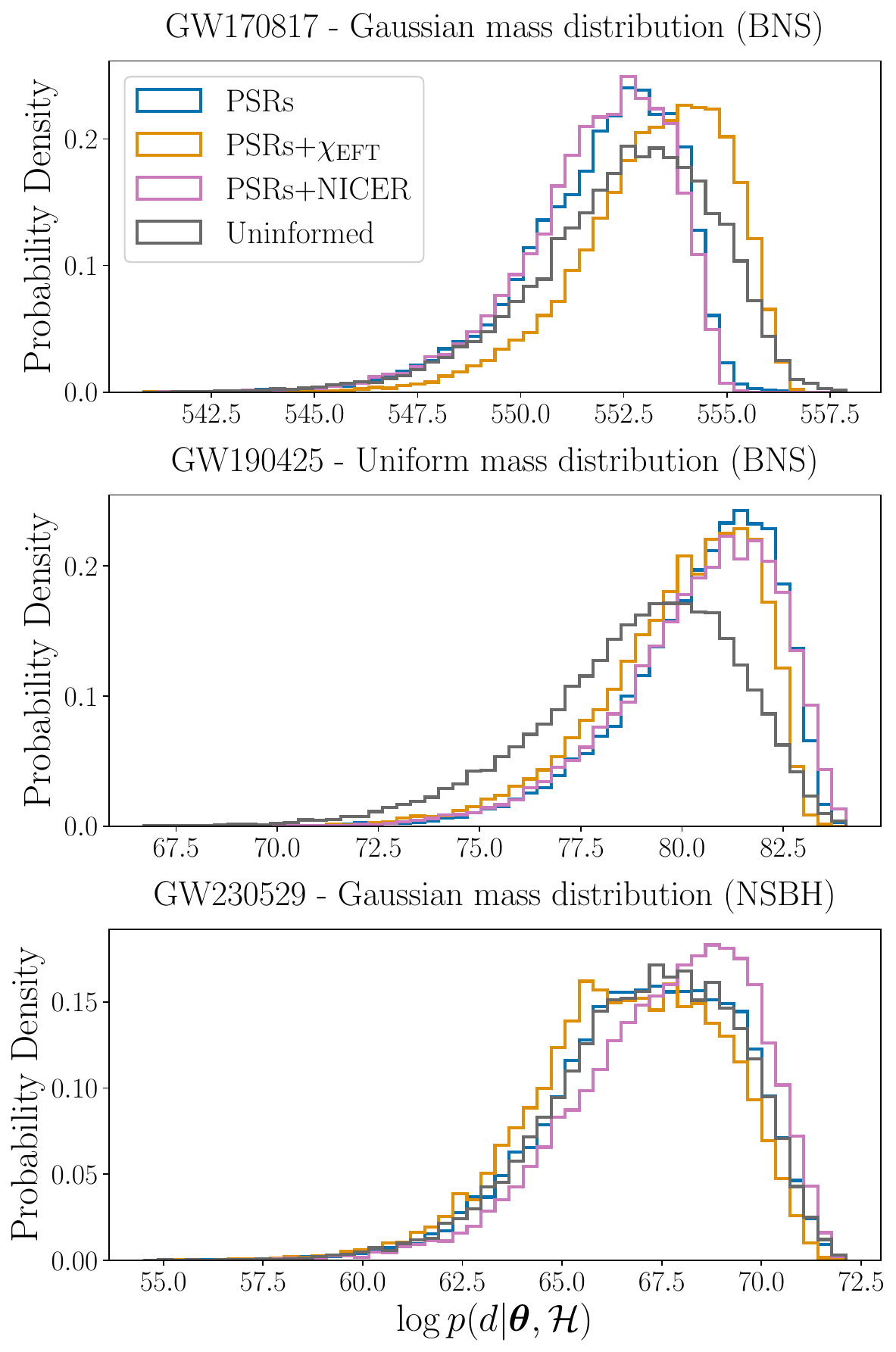}
    \caption{Histograms of the log-likelihood values of the posterior samples obtained with the uninformed prior (gray) and the physics-informed priors (colors).
    For each \ac{GW} event, we only show the results for the preferred \ac{NS} mass distribution.
    }
    \label{fig: combined log likelihood plots}
\end{figure}

As a result, the histograms of log-likelihood values overlap quite well between the different priors, as shown in the bottom panel of Fig.~\ref{fig: combined log likelihood plots}.
The physics-informed prior that additionally uses information from $\chiEFT$, respectively \ac{NICER}, leads to a lower, respectively higher, median value, although the changes are too small to draw quantitative conclusions.

{
\renewcommand{\arraystretch}{1.25}
\begin{table*}[t]
    \centering
    \caption{Bayes factors in $\log_{10}$ scale between the hypothesis obtaining the highest Bayesian evidence for a \ac{GW} event (marked `ref.') and other hypotheses.
    The entries of the table are color-coded according to their interpretation of the Jeffrey's scale (see Table~\ref{tab:jeffreys_scale}), with fainter (darker) shades of red referring to weaker (stronger) evidence.
    }
    \label{tab: Bayes factors}
    \begin{tabular}{|l|l|l|c|c|c|}
\hline
\textbf{Source} & \textbf{Mass distribution} & \textbf{EOS Constraints} & \textbf{GW170817} & \textbf{GW190425} & \textbf{GW230529} \\
\hline\hline
\multirow{9}{*}{BNS} & \multirow{3}{*}{Uniform} & PSRs & \cellcolor{jeffreysred4}$-1.83$ & \cellcolor{jeffreysred1}$-0.07$ & \cellcolor{jeffreysred5}$-13.14$ \\
 &  & PSRs+$\chi_{\rm{EFT}}$ & \cellcolor{jeffreysred2}$-0.80$ & \cellcolor{jeffreysred1}$-0.11$ & \cellcolor{jeffreysred5}$-13.12$ \\
 &  & PSRs+NICER & \cellcolor{jeffreysred4}$-1.58$ & \textbf{ref.} & \cellcolor{jeffreysred5}$-12.92$ \\
\cline{2-6}
 & \multirow{3}{*}{Gaussian} & PSRs & \cellcolor{jeffreysred2}$-0.68$ & \cellcolor{jeffreysred5}$-6.89$ & \cellcolor{jeffreysred5}$-18.82$ \\
 &  & PSRs+$\chi_{\rm{EFT}}$ & \textbf{ref.} & \cellcolor{jeffreysred5}$-8.47$ & \cellcolor{jeffreysred5}$-18.83$ \\
 &  & PSRs+NICER & \cellcolor{jeffreysred2}$-0.76$ & \cellcolor{jeffreysred5}$-5.45$ & \cellcolor{jeffreysred5}$-18.81$ \\
\cline{2-6}
 & \multirow{3}{*}{Double Gaussian} & PSRs & \cellcolor{jeffreysred3}$-1.36$ & \cellcolor{jeffreysred2}$-0.55$ & \cellcolor{jeffreysred5}$-13.75$ \\
 &  & PSRs+$\chi_{\rm{EFT}}$ & \cellcolor{jeffreysred2}$-0.59$ & \cellcolor{jeffreysred2}$-0.79$ & \cellcolor{jeffreysred5}$-13.77$ \\
 &  & PSRs+NICER & \cellcolor{jeffreysred2}$-0.92$ & \cellcolor{jeffreysred2}$-0.57$ & \cellcolor{jeffreysred5}$-13.71$ \\
\cline{2-6}
\hline\hline
\multirow{9}{*}{NSBH} & \multirow{3}{*}{Uniform} & PSRs & \cellcolor{jeffreysred5}$-224.65$ & \cellcolor{jeffreysred4}$-1.52$ & \cellcolor{jeffreysred1}$-0.08$ \\
 &  & PSRs+$\chi_{\rm{EFT}}$ & \cellcolor{jeffreysred5}$-224.66$ & \cellcolor{jeffreysred3}$-1.35$ & \cellcolor{jeffreysred1}$-0.02$ \\
 &  & PSRs+NICER & \cellcolor{jeffreysred5}$-224.66$ & \cellcolor{jeffreysred4}$-1.63$ & \cellcolor{jeffreysred1}$-0.25$ \\
\cline{2-6}
 & \multirow{3}{*}{Gaussian} & PSRs & \cellcolor{jeffreysred5}$-224.67$ & \cellcolor{jeffreysred2}$-0.82$ & \cellcolor{jeffreysred1}$-0.05$ \\
 &  & PSRs+$\chi_{\rm{EFT}}$ & \cellcolor{jeffreysred5}$-224.66$ & \cellcolor{jeffreysred3}$-1.11$ & \cellcolor{jeffreysred1}$-0.20$ \\
 &  & PSRs+NICER & \cellcolor{jeffreysred5}$-224.66$ & \cellcolor{jeffreysred3}$-1.43$ & \textbf{ref.} \\
\cline{2-6}
 & \multirow{3}{*}{Double Gaussian} & PSRs & \cellcolor{jeffreysred5}$-224.67$ & \cellcolor{jeffreysred5}$-4.11$ & \cellcolor{jeffreysred1}$-0.14$ \\
 &  & PSRs+$\chi_{\rm{EFT}}$ & \cellcolor{jeffreysred5}$-224.68$ & \cellcolor{jeffreysred5}$-3.83$ & \cellcolor{jeffreysred1}$-0.13$ \\
 &  & PSRs+NICER & \cellcolor{jeffreysred5}$-224.67$ & \cellcolor{jeffreysred5}$-24.31$ & \cellcolor{jeffreysred1}$-0.05$ \\
\cline{2-6}
\hline
\end{tabular}
\end{table*}
}

\subsection{Parameter constraints}\label{sec: results: parameter constraints}

In the following subsections, we discuss the posteriors obtained from the physics-informed priors detailed in Sec.~\ref{sec: methods: neural priors}, limiting ourselves to the mass distribution to which the reference model belongs.
We focus the discussion on selected \ac{GW} parameters which show remarkable differences compared to the uninformed priors.
The constraints are also given as $90\%$ credible intervals in Table~\ref{tab: parameter values} in Appendix~\ref{sec: app: parameter values}.

\subsubsection{GW170817}\label{sec: results: parameter constraints: GW170817}

Figure~\ref{fig: GW170817 inference results} shows the posterior distributions for the chirp mass $\mathcal{M}_c$, the mass ratio $q$, the mass-weighted tidal deformabilities $\LambdaTilde$ and $\DeltaLambdaTilde$, and the luminosity distance $d_L$ for the GW170817 signal.
Regarding the masses, we note that the posteriors on the chirp mass have reduced uncertainties when using the physics-informed priors compared to the uninformed prior.
The posteriors for mass ratio overlap significantly, although small differences are noticeable due to the correlations with $\DeltaLambdaTilde$ that are present in the priors (see Fig.~\ref{fig: neural priors}).
While the informed posteriors have most of their probability mass lying above $q = 0.9$, they have negligible support for an equal mass system (i.e., $q = 1$), in contrast to the agnostic prior.
This feature is due to the Gaussian mass distribution prior.
Indeed, we have verified that the mass ratio posterior shows a similar feature when sampling the source-frame masses from the Gaussian mass distribution while instead sampling $\Lambda_i \in [0, 5000]$ rather than informed by the \ac{EOS}, which gives $q = 0.95_{-0.03}^{+0.03}$ at the $90\%$ credible level.

As for the tidal deformabilities, the physics-informed priors recover narrower posteriors on $\LambdaTilde$.
As expected, the posterior values reflect the \ac{EOS} constraints embedded into the prior, as the median is lower for the prior using information from $\chiEFT$ compared to the other physics-informed priors.
Our reference model (using the PSRs+$\chiEFT$ constraint) infers $\LambdaTilde = 385_{-85}^{+141}$ ($90\%$ credibility).
Since $\DeltaLambdaTilde$ has a less dominant contribution to the \ac{GW} phase than $\LambdaTilde$, there is less distinction in its posterior between the different hypotheses.

Finally, we note that the posterior on the luminosity distance is shifted to higher median values when using the physics-informed priors compared to the uninformed prior, as also apparent from Table~\ref{tab: parameter values}.
This is because the source-frame chirp mass is constrained to lower values.
Therefore, the redshift (or equivalently, the luminosity distance) must be increased to be consistent with the measured detector-frame chirp mass.

Interestingly, we note that the posteriors we derived are consistent with inferences that use information from the \ac{EM} counterpart of GW170817~\cite{LIGOScientific:2017ync}.
For instance, inferences that jointly model the \ac{GW} and kilonova signal prefer more equal mass systems~\cite{Radice:2018ozg, Coughlin:2018fis, Pang:2022rzc, Breschi:2024qlc}.
Moreover, inference of the kilonova signal also leads to the constraint of $\LambdaTilde \gtrsim 300$~\cite{Radice:2018ozg, Coughlin:2018fis, Pang:2022rzc, Breschi:2024qlc}.
Finally, fixing the sky location to that of the host galaxy excludes lower luminosity distances~\cite{LIGOScientific:2018hze}.
However, our work establishes these constraints purely based on analysis of the \ac{GW} data and without invoking the \ac{EM} information at any point.

{
\begin{figure}[t]
    \centering
    \includegraphics[width=\columnwidth]{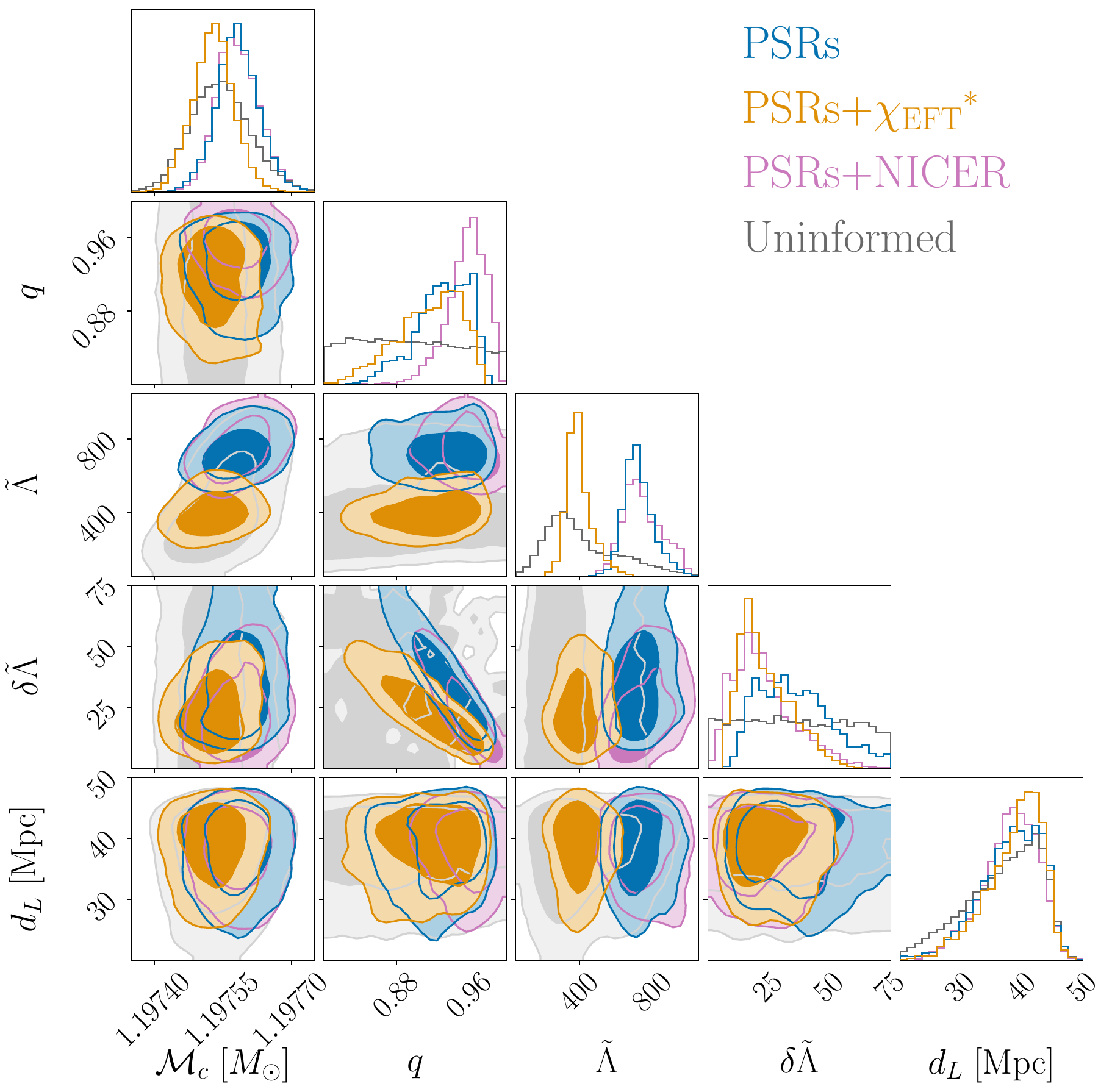}
    \caption{Inference results of some selected parameters for GW170817 using the \ac{BNS} hypothesis with the Gaussian mass distribution and varying \ac{EOS} constraints (colors) compared to the inference using uninformed priors (in gray).
    The star indicates the reference model with the highest evidence across the different physics-informed priors.
    The dark (light) contours denote the $68\%$ ($95\%$) credible area.
    }
    \label{fig: GW170817 inference results}
\end{figure}
}

\subsubsection{GW190425}\label{sec: results: parameter constraints: GW190425}

\begin{figure}[t]
    \centering
    \includegraphics[width=\columnwidth]{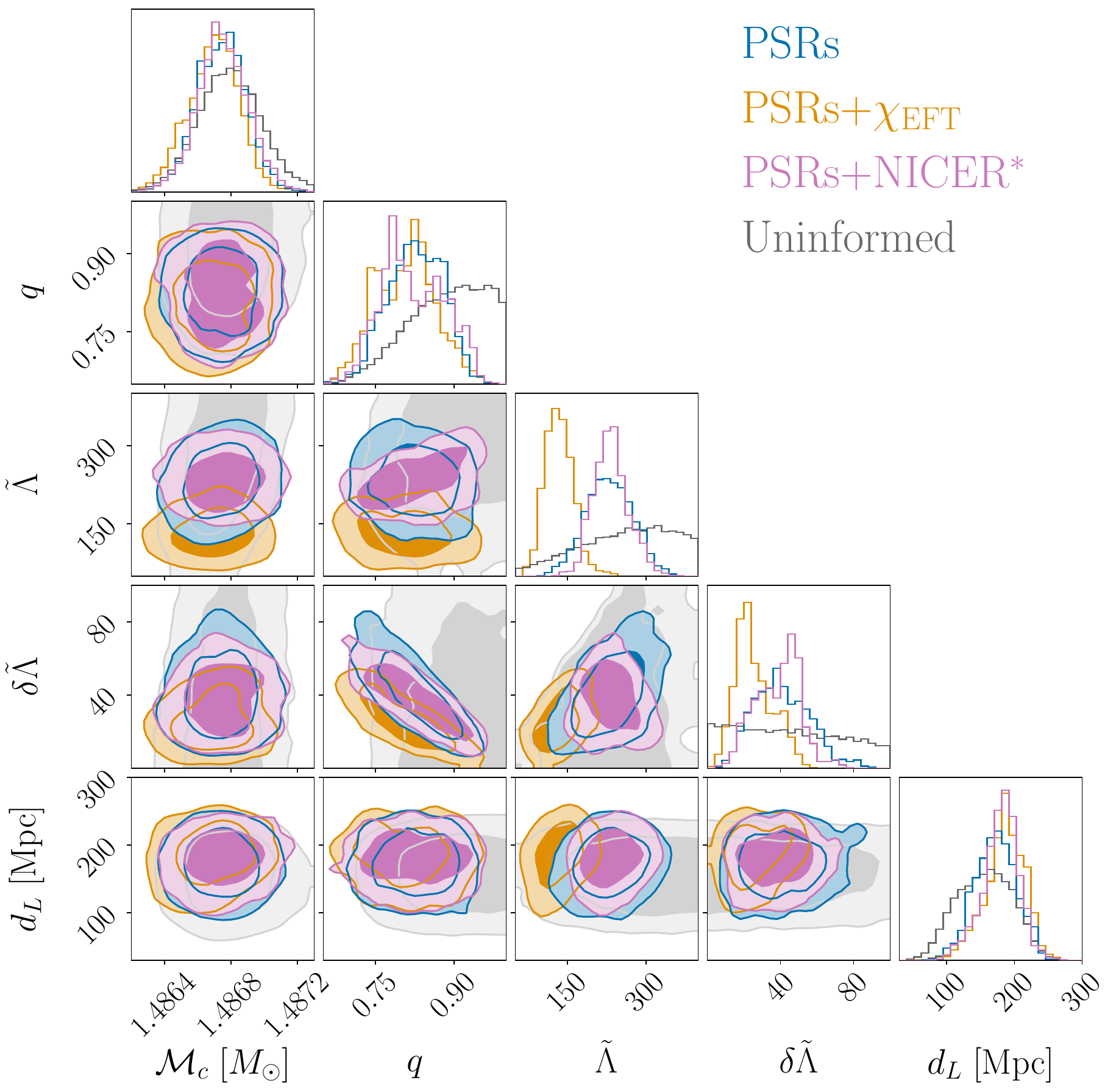}
    \caption{Inference results of some selected parameters for GW190425 using the \ac{BNS} hypothesis with the uniform mass distribution and varying \ac{EOS} constraints (colors) compared to the inference using uninformed priors (in gray).
    The star indicates the reference model with the highest evidence across the different physics-informed priors.
    The dark (light) contours denote the $68\%$ ($95\%$) credible area.
    }
    \label{fig: GW190425 inference results}
\end{figure}

Figure~\ref{fig: GW190425 inference results} shows the posterior distributions for the source parameters of GW190425.
Concerning the masses, the posteriors on the chirp mass show more overlap between the different priors compared to the GW170817 analysis, which can be attributed to the lower \ac{SNR} of the signal.
For the mass ratio, the physics-informed priors consistently recover systems with less equal masses compared to the uninformed priors, having negligible support for mass ratios above $0.9$.
This preference for a lower mass ratio is not present in the priors (see Fig.~\ref{fig: neural priors}), and therefore must come from the signal and correlations between parameters introduced by the physics-informed priors.

The posteriors on $\LambdaTilde$ cluster around values consistent with our prior expectations.
As for GW170817, the median value is shifted based on the stiffness of the \ac{EOS} constraints used in the physics-informed prior, whereas the posteriors for $\DeltaLambdaTilde$ overlap quite well.

The inferred luminosity distance increases when using the physics-informed priors, as we also observed for GW170817.
However, compared to GW170817, the difference between the informed and agnostic inferences is more apparent, likely due to the larger distance at which GW190425 occurred.
As shown in Table~\ref{tab: parameter values}, our reference model recovers a luminosity distance of $182^{+41}_{-49}$ Mpc, with other physics-informed priors leading to similar values.
The agnostic inference, on the other hand, recovers a distance $157^{+64}_{-65}$ Mpc (both at $90\%$ credibility).

\subsubsection{GW230529}\label{sec: results: parameter constraints: GW230529}

\begin{figure}[t]
    \centering
    \includegraphics[width=\columnwidth]{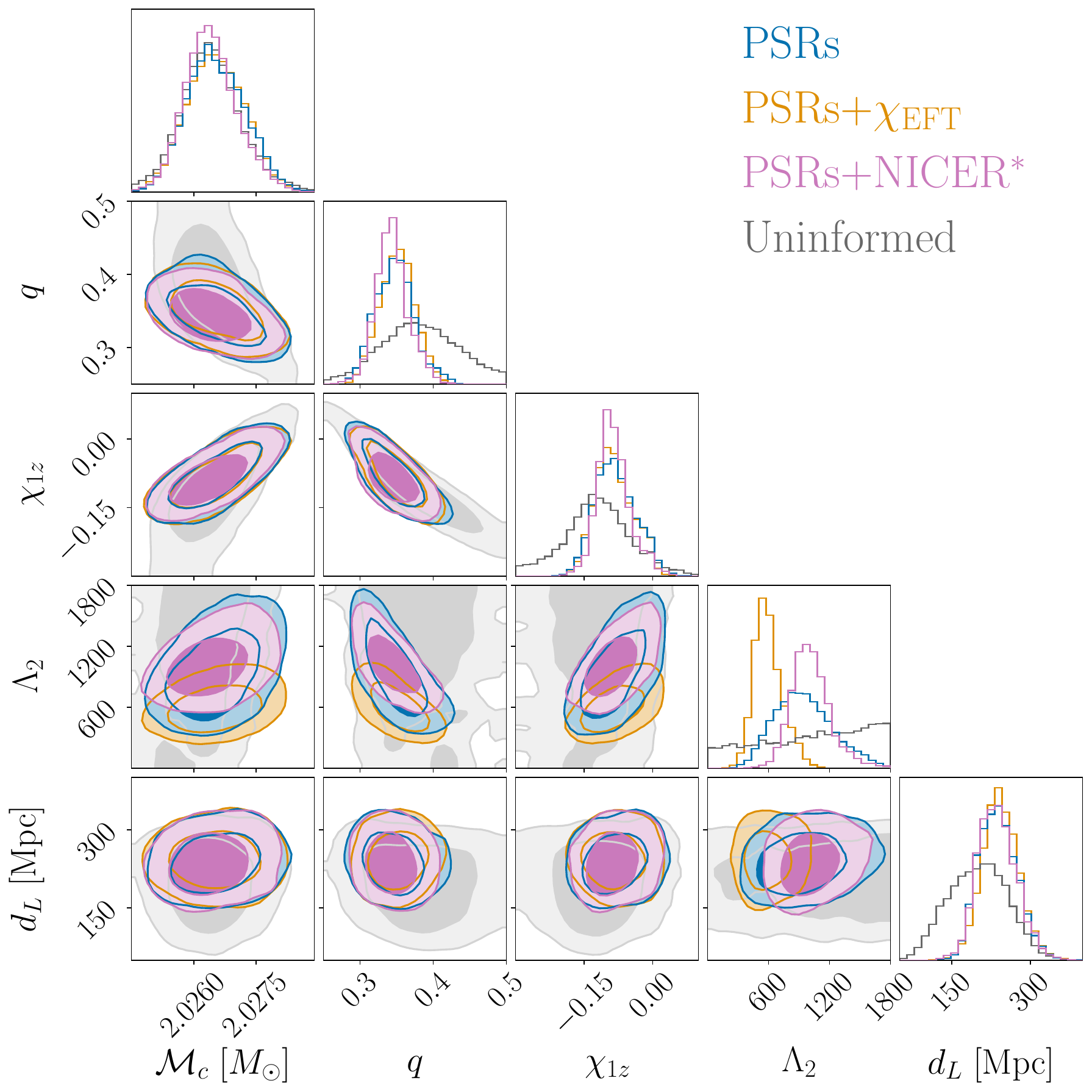}
    \caption{Inference results of some selected parameters for GW230529 using the \ac{NSBH} hypothesis with the Gaussian mass distribution and varying \ac{EOS} constraints (colors) compared to the inference using uninformed priors (in gray).
    The star indicates the reference model with the highest evidence across the different physics-informed priors.
    The dark (light) contours denote the $68\%$ ($95\%$) credible area.
    }
    \label{fig: GW230529 inference results}
\end{figure}

Figure~\ref{fig: GW230529 inference results} shows the posteriors for source parameters of GW230529, namely, the chirp mass, mass ratio, $z$-component of the spin of the primary $\chi_{1z}$, the tidal deformability for the secondary component $\Lambda_2$, and the luminosity distance.
Due to the low \ac{SNR} of the signal, we find significant overlap between the different physics-informed priors, consistent with the observation that the Bayes factors between them are quite low (see Table~\ref{tab: Bayes factors}).

Regarding the masses, the physics-informed priors recover narrower posteriors on the mass ratio, excluding the tail above $q\geq 0.4$ ($90\%$ credibility).
Due to a degeneracy between the mass ratio and spins~\cite{Baird:2012cu, Ng:2018neg}, this leads to improved measurements on the spin parameters of the primary component.
In particular, the physics-informed priors yield values for $\chi_{1z}$ that are closer to zero compared to the uninformed inference.
Similar constraints on the mass ratio and spins were also obtained by Ref.~\cite{Chattopadhyay:2024hsf}.
However, Ref.~\cite{Chattopadhyay:2024hsf} used astrophysically informed priors on masses and spins, whereas we employed uninformed spin priors.

The tidal deformability posteriors are more prior-dominated for GW230529.
As already explained in Sec.~\ref{sec: results: model selection: GW230529}, this is due to the amplitude tapering of the \texttt{NRTidalv3} waveform, which limits our sensitivity to tidal deformations in the GW230529 signal.

Finally, we find that the physics-informed priors result in a higher luminosity distance compared to the uninformed prior, as was the case for GW170817 and GW190425.
Our reference model finds a luminosity distance of $235_{-58}^{+59}$ Mpc, compared to $201_{-97}^{+84}$ Mpc for the uninformed inference (both at $90\%$ credibility).
While detailed calculations regarding the possibility of an \ac{EM} counterpart of the merger are beyond the scope of this work (but see, e.g., Refs.~\cite{Zhu:2024cvt, Chandra:2024ila, Martineau:2024zur, Kunnumkai:2024qmw, Markin:2025oeo}), this makes it less likely that follow-up observations would have been able to detect a potential \ac{EM} counterpart~\cite{Pillas:2025pfc}.

\section{Discussion}\label{sec: discussion}

\subsection{Source classification}

Inferring the nature of compact objects observed in \ac{GW} events from their inspiral is challenging for objects with masses close to the \ac{TOV} mass~\cite{Hannam:2013uu, Littenberg:2015tpa}.
One approach relies on the mass estimates derived from \ac{GW} inferences.
For instance, by computing the overlap between an object's mass posterior and the posterior on $\MTOV$, the probability that the object is a \ac{BH} or an \ac{NS} can be estimated~\cite{Essick:2019ldf, Fishbach:2020ryj, Essick:2020ghc, LIGOScientific:2020zkf, Tews:2020ylw, Tsokaros:2020hli, Farah:2021qom, LIGOScientific:2024elc, Koehn:2024ape, Mali:2025etk}.
In particular, Ref.~\cite{Essick:2020ghc} used an explicit mixture model, where classification is performed concurrently within hierarchical inference, rather than as a separate single-event inference before a hierarchical inference, as done in this work.
However, this implicitly assumes that the \ac{NS} and \ac{BH} populations are non-overlapping~\cite{Fishbach:2020ryj}, ignoring the possibility that compact objects in the \ac{NS} mass regime could be (primordial) \acp{BH}~\cite{Kouvaris:2018wnh, Carr:2020xqk, Carr:2020gox, Dasgupta:2020mqg, Clesse:2020ghq, Tsai:2020hpi, Singh:2022wvw} or exotic compact objects~\cite{Mazur:2004fk, Colpi:1986ye, Liebling:2012fv}.

Besides the masses, one can hope to identify compact objects through their tidal signature~\cite{Cardoso:2017cfl, Sennett:2017etc, Johnson-Mcdaniel:2018cdu}.
In practice, however, the difference in the waveforms due to tidal deformations is small~\cite{Matas:2020wab}, such that this approach only works well for sufficiently loud signals such as observed by future ground-based detectors~\cite{Chen:2020fzm, Brown:2021seh, Thompson:2020nei,  Franciolini:2021xbq, Datta:2020gem, Cotturone:2025jlm, Khadkikar:2025awf, Bhattacharya:2025xko, Bhattacharya:2023stq}, for low-mass \acp{NS} with large tidal deformabilities~\cite{Crescimbeni:2024cwh, Golomb:2024mmt, Coupechoux:2021bla}, or after observing a population of mergers~\cite{Fasano:2020eum, Muller:2025vwr}.

In this work, we presented a method for Bayesian source classification under the assumption that \acp{NS} and \acp{BH} have non-overlapping mass distributions.
In future work, we will explore how well this approach can distinguish compact objects when this assumption is relaxed, relying on physics-informed priors to constrain the relation between mass and tidal deformability \textit{a priori}.

\subsection{Prior to posterior information gain}

\begin{figure}[t]
    \centering
    \includegraphics[width=0.99\columnwidth]{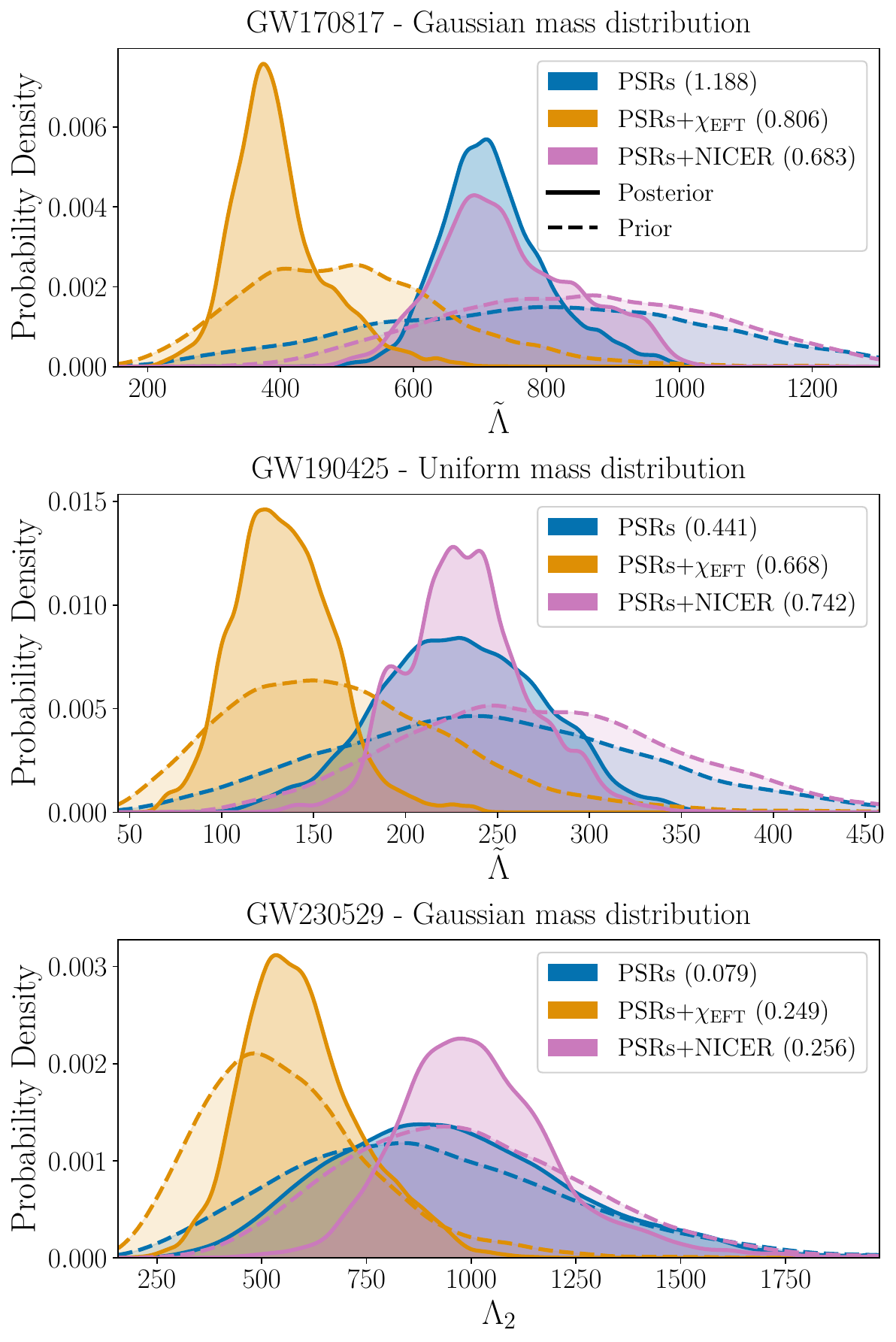}
    \caption{Comparison between the priors (dashed lines) on tidal deformability, conditioned on the posterior source-frame masses, and the corresponding posteriors (solid lines).
    The values in brackets denote the \ac{KL} divergence in bits between each prior and posterior pair.
    }
    \label{fig: prior vs posterior}
\end{figure}

Since the physics-informed priors, by construction, restrict the parameter space explored during inference, it is important to determine to what extent the posteriors in Sec.~\ref{sec: results: parameter constraints} are shaped by the prior or by the information provided by the signal.
To assess this, we compute the information gain between the prior and posterior distributions for the tidal deformability.
We construct the priors by sampling the \acp{EOS} from the sets discussed in Sec.~\ref{sec: methods: neural priors: EOS} and evaluating their predictions for the tidal deformabilities at the source-frame component masses of the posterior samples.
We quantify the degree of similarity between these priors and the posteriors through the \ac{KL} divergence~\cite{Boyd:2004fnq}.
The comparison is shown visually in Fig.~\ref{fig: prior vs posterior}, and the \ac{KL} divergences (measured in bits) between prior and posterior pairs are given in the brackets in the legend.
We note that, as expected, the \ac{KL} divergence is larger if the \ac{SNR} of the signal is larger, as there is more information content in the signal driving the inference.
Nevertheless, we still find noticeable shifts in the posteriors due to the signal in other events and physics-informed priors, showing that the analyses are not simply reproducing the informed priors.

\subsection{Equation-of-state-informed GW inference}

Previous works have considered different methods to use \ac{EOS} information during \ac{GW} parameter estimation for \ac{BNS} systems.
One example is the use of universal relations~\cite{Yagi:2015pkc, LIGOScientific:2018cki, Chatziioannou:2018vzf}.
However, these relations have residuals that have to be accounted for~\cite{Chatziioannou:2018vzf}, and can lead to conceptual or technical problems when used in Bayesian inference~\cite{Kastaun:2019bxo}.
Moreover, these relations are designed without assuming any constraints on the \ac{EOS}, and would therefore need to be modified to explicitly incorporate such information.

Instead, the \ac{EOS} can be sampled on the fly in two ways.
First, one can choose a parametrization for the \ac{EOS} and jointly sample the \ac{GW} and \ac{EOS} parameters, using machine learning emulators of the \ac{TOV} solver to accelerate the likelihood evaluation~\cite{Magnall:2024ffd, Reed:2024urq, Somasundaram:2024ykk, Reed:2025sqh}.
However, this leads to an increase in the number of parameters to sample, which makes convergence less efficient.
Second, one can use a discrete set of tabulated \acp{EOS}, which is either sampled during \ac{GW} inference~\cite{LIGOScientific:2019eut, Pang:2022rzc, Rose:2023uui, Vilkha:2024hgt, Koehn:2024set, Wouters:2025ull}, or used in the postprocessing of \ac{GW} posteriors that relied on agnostic priors~\cite{Ghosh:2021eqv, Biswas:2021pvm, Khadkikar:2025ith, Kashyap:2025cpd}.
However, this requires more memory to store the tabulated \acp{EOS} and might lead to a low number of posterior samples when used to analyze high \ac{SNR} signals~\cite{Rose:2023uui}.

Finally, another option is to emulate the distributions of the relevant parameters for \ac{GW} inference as predicted by the \ac{EOS} constraints, which alleviates the issues described above.
To this end, Ref.~\cite{Magnall:2025zhm} recently introduced physics-informed priors based on \ac{EOS} information for \ac{GW} inference, making it conceptually similar to the present study.
However, their method was limited to modeling the conditional distribution of $\LambdaTilde$ given the source-frame chirp mass.
Our approach instead models the joint distributions of both component masses and tidal deformabilities, offering a more general framework to directly incorporate \ac{EOS} constraints into \ac{GW} inference.

\subsection{Future work}

Before concluding, we highlight a few of our assumptions and how these present themselves as opportunities to extend the methodology in future work.

First, although we have not followed the mathematically stricter approach of hierachichal inference to incorporate information about the \ac{NS} mass distributions and the \ac{EOS} directly from the \ac{GW} data itself, our method also allows to be extended to include the latest population models released by the \ac{LVK} collaboration~\cite{LIGOScientific:2025pvj} or more \ac{EOS} constraints, including the multimessenger inferences of GW170817~\cite{Radice:2017lry, Capano:2019eae, Radice:2018ozg, Dietrich:2020efo, Pang:2022rzc}.

Second, we note that this work focused on the application of physics-informed priors in \ac{GW} inference.
For multimessenger events like GW170817, joint inference on all messengers could further constrain parameters and aid classification.
However, when using agnostic priors in such a joint analysis, the Bayes factors still favor the \ac{BNS} hypothesis only mildly~\cite{Coughlin:2019kqf}.
In the future, we will extend our physics-informed priors to additionally model parameters of the \ac{EM} counterpart.
For instance, by making use of the fits in Refs.~\cite{Agathos:2019sah, Kruger:2020gig, Dietrich:2020efo, Breschi:2024qlc}, the \ac{GW} binary parameters can be linked to the mass ejecta of kilonovae~\cite{Pang:2022rzc, Breschi:2024qlc} (although the fits might be subject to systematic errors~\cite{Ristic:2025bvt}).
In this way, the physics-informed priors can be used for informed multimessenger analyses.

Third, we note that we have neglected phenomena that can soften or stiffen the \ac{EOS}, such as dark matter~\cite{Panotopoulos:2017idn, Das:2018frc, Ellis:2018bkr, Nelson:2018xtr, Quddus:2019ghy}, pressure anisotropy~\cite{Herrera:1997plx, Herrera:2020gdg, Pang:2025fes}, phase transitions~\cite{Paschalidis:2017qmb, Nandi:2017rhy, Han:2018mtj, Raithel:2022aee}, modelling of the nuclear physics~\cite{Rose:2023uui, Legred:2025aar, Shakeri:2022dwg, Shahrbaf:2024gdm}, or various compositions~\cite{Sun:2018tmw, Li:2019tjx, Ribes:2019kno, Han:2019bub, Ferreira:2020evu, Blaschke:2020qqj}.
Therefore, the preference for soft \acp{EOS} could arise from (a combination of) the aforementioned effects rather than a purely hadronic, soft \ac{EOS}.
However, the physics-informed priors, being data-driven, can perfectly accommodate any set of \acp{EOS} in the construction (see Sec.~\ref{sec: methods: neural priors construction details}).
Future work will investigate whether the model selection introduced in this work can generalize to discern such effects in sufficiently loud \ac{BNS} mergers or a collection of \ac{BNS} mergers.

Finally, future work will consider, with a simulation study, whether the use of informed priors in \ac{GW} inference results in improved constraints on the \ac{EOS}~\cite{Magnall:2025zhm}.

\section{Conclusion}\label{sec: conclusion}

Data analysis of \ac{GW} signals from mergers involving \acp{NS} typically employs agnostic priors on the source parameters.
In this work, we propose a flexible, data-driven method to incorporate external knowledge of \ac{NS} physics into \ac{GW} parameter estimation.
In particular, we train \acp{NF}, a type of neural density estimator, on predictions of masses and tidal deformabilities for \ac{BNS} and \ac{NSBH} systems from \ac{NS} mass distributions and existing constraints on the \ac{EOS}.

We construct such priors informed by existing \ac{EOS} constraints and mass distributions.
Applying them to inferences of GW170817, GW190425, and GW230529, we highlight two benefits from using the physics-informed priors over agnostic priors.
First, we demonstrate that we can perform Bayesian model selection by computing Bayes factors between hypotheses encoded as different physics-informed priors.
We classify GW170817 as a \ac{BNS} merger, and find substantial evidence for a soft \ac{EOS} to explain the tidal deformation in the signal.
GW190425 is classified as a \ac{BNS} merger, with strong to decisive evidence, while GW230529 is decisively characterized as an \ac{NSBH} merger.
While these findings are in line with previous works, we assess them based on data-driven priors directly from analyzing the \ac{GW} data.

Second, the physics-informed priors result in narrower posteriors on the \ac{GW} source parameters.
We find that the informed constraints derived for GW170817 from the \ac{GW} data alone are consistent with multimessenger analyses of the merger that jointly model the \ac{GW} data and \ac{EM} counterpart.
In particular, we find a preference for nearly-equal mass systems, constrain $\LambdaTilde\gtrsim 300$, and recover a luminosity distance consistent with information from the host galaxy.
For GW190425, our informed posteriors favor less equal masses compared to agnostic priors, with almost no posterior support for mass ratios above $0.9$.
Regarding GW230529, we find narrower constraints on the mass ratio and therefore the spins of the binary, while the tidal deformation posterior is more prior-dominated due to the termination of the waveform model at low frequencies.
For GW190425 and GW230529, we recover a significantly larger luminosity distance compared to agnostic inferences, due to a more constrained source-frame chirp mass.

Looking ahead, we envision that our data-driven methodology can easily incorporate information from additional sources and can be generalized to inform other parameters of interest.
Therefore, it can easily accommodate the advances in our understanding of both the \ac{NS} population and dense nuclear matter in the coming years.

\section*{Data availability}

The physics-informed priors are implemented directly in the \textsc{bilby} source code available at \url{https://github.com/ThibeauWouters/bilby/tree/neural_prior_bilby_pipe}~\cite{bilby_src_code}.
Accompanying code and data used to produce the results shown in this work are available at \url{https://github.com/ThibeauWouters/neural_priors/}~\cite{paper_repo}.

\section*{Acknowledgments}

We thank Harsh Narola, Ann-Kristin Malz, John Veitch, Eline van Straaten, Michael Williams, Lami Suleiman, Sanika Khadkikar, and the LIGO-Virgo KAGRA parameter estimation and extreme matter working groups for their fruitful discussions and feedback, which led to the improvement of this work.
T.W. and C.V.D.B. are supported by the research program of the Netherlands Organization for Scientific Research (NWO) through grant number OCENW.XL21.XL21.038.
P.T.H.P. is supported by the research
program of the Netherlands Organization for Scientific
Research (NWO) under grant number VI.Veni.232.021.
T.D. acknowledges funding from the EU Horizon under ERC Starting Grant, no. SMArt-101076369.
We thank SURF (www.surf.nl) for the support in using the National Supercomputer Snellius under project numbers EINF-14622.
Some computations were performed on the DFG-funded research cluster Jarvis at the University of Potsdam (INST 336/173-1; project number: 502227537).
Views and opinions expressed are those of the authors only and do not necessarily reflect those of the European Union or the European Research Council. Neither the European Union nor the granting authority can be held responsible for them. This research has made use of data or software obtained from the Gravitational Wave Open Science Center (gwosc.org), a service of the LIGO Scientific Collaboration, the Virgo Collaboration, and KAGRA. This material is based upon work supported by NSF's LIGO Laboratory which is a major facility fully funded by the National Science Foundation, as well as the Science and Technology Facilities Council (STFC) of the United Kingdom, the Max-Planck-Society (MPS), and the State of Niedersachsen/Germany for support of the construction of Advanced LIGO and construction and operation of the GEO600 detector. Additional support for Advanced LIGO was provided by the Australian Research Council. Virgo is funded, through the European Gravitational Observatory (EGO), by the French Centre National de Recherche Scientifique (CNRS), the Italian Istituto Nazionale di Fisica Nucleare (INFN) and the Dutch Nikhef, with contributions by institutions from Belgium, Germany, Greece, Hungary, Ireland, Japan, Monaco, Poland, Portugal, Spain. KAGRA is supported by Ministry of Education, Culture, Sports, Science and Technology (MEXT), Japan Society for the Promotion of Science (JSPS) in Japan; National Research Foundation (NRF) and Ministry of Science and ICT (MSIT) in Korea; Academia Sinica (AS) and National Science and Technology Council (NSTC) in Taiwan.
T.W. acknowledges the use of generative AI (Claude Sonnet 4.5) to proofread the manuscript.
All AI-generated outputs were carefully reviewed and edited by T.W. to ensure accuracy.

\appendix

\section{Gravitational wave prior distributions}\label{sec: app: GW parameters and priors}

The \texttt{IMRPhenomXP\_NRTidalv3} waveform has $17$ parameters in total.
Their default priors are listed in Table~\ref{tab: GW parameters and priors}~\cite{Romero-Shaw:2020owr}.
The default prior on the masses samples the detector-frame component masses uniformly while restricting the chirp mass and mass ratio $q$ to particular ranges.
In Table~\ref{tab: GW parameters and priors}, this is denoted by the symbol $\mathcal{U}_{\comp}$, with the ranges denoted in the brackets.
For \acp{BH}, spin magnitudes are varied up to the maximum value of $0.99$, while for \acp{NS}, we restrict this range to be below $0.05$.
The geocentric time is sampled in a uniform interval of width $0.1$ centered on the \ac{GW} trigger time for each event given in the table.
For the \ac{NSBH} hypothesis, we sample $\Lambda_2$ in the interval $[0, 5000]$ and fix $\Lambda_1 = 0$.

\renewcommand{\arraystretch}{1.25}
\begin{table}[t]
\centering
\caption{
    Table of \ac{GW} parameters and their prior distributions.
    $\mathcal{U}$ refers to a uniform distribution, $\mathcal{U}_{\rm{comp}}$ refers to a mass distribution that is uniform in detector-frame masses while restricting the parameter to the range given by the brackets, and $\mathcal{U}_{\rm{com. vol}}$ refers to a luminosity distance distribution that is uniform in comoving volume.
    }
\label{tab: GW parameters and priors}
\begin{tabular}{l|llcc}
\hline\hline
& & parameter & symbol & prior \\
\hline
\multirow{10}{*}{\rotatebox{90}{common}} & & Spin magnitude NS & $a_j$ & $\mathcal{U}(0, 0.05)$ \\
& & Spin magnitude BH & $a_j$ & $\mathcal{U}(0, 0.99)$ \\
& & Tilt & $\sin \varphi_j$ & $\mathcal{U}(-1, 1)$ \\
& & Misalignment [rad] & $\phi_{12}, \phi_{JL}$ & $\mathcal{U}(0, 2\pi)$ \\
& & Right ascension [rad] & $\alpha$ & $\mathcal{U}(0, 2\pi)$ \\
& & Declination [rad] & $\delta$ & $\mathcal{U}(0, 2\pi)$ \\
& & Inclination & $\theta_{JN}$ & Cosine$(0, \pi)$ \\
& & Phase [rad] & $\phi$ & $\mathcal{U}(0, 2\pi)$ \\
& & Polarization [rad] & $\psi$ & $\mathcal{U}(0, 2\pi)$ \\
& & Time delay [s] & $\delta t_c$ & $\mathcal{U}(-0.1, 0.1)$ \\
\hline
\multirow{3}{*}{\rotatebox{90}{phys.-inf.}} & & Source-frame chirp mass & $\mathcal{M}_c^\src$ &  \\
& & Mass ratio & $q$ & see Sec.~\ref{sec: methods: neural priors} \\
& & Tidal deformabilities & $\Lambda_i$ &  \\
\hline
\multirow{15}{*}{\rotatebox{90}{uninformed}} & \multirow{5}{*}{\rotatebox{90}{GW170817}} & Chirp mass [M$_\odot$] & $\mathcal{M}_c$ & $\mathcal{U}_{\rm{comp}}(1.18, 1.21)$ \\
& & Mass ratio & $q$ & $\mathcal{U}_{\rm{comp}}(0.125, 1)$ \\
& & Tidal deformability & $\Lambda_i$ & $\mathcal{U}(0, 5000)$ \\
& & Luminosity distance [Mpc] & $d_L$ & $\mathcal{U}_{\text{com. vol.}}(1, 75)$ \\
& & Trigger time [GPS] & $t_c$ & $1187008882.43$ \\
\cline{2-5}
& \multirow{5}{*}{\rotatebox{90}{GW190425}} & Chirp mass [M$_\odot$] & $\mathcal{M}_c$ & $\mathcal{U}_{\rm{comp}}(1.485, 1.49)$ \\
& & Mass ratio & $q$ & $\mathcal{U}_{\rm{comp}}(0.125, 1)$ \\
& & Tidal deformability & $\Lambda_i$ & $\mathcal{U}(0, 5000)$ \\
& & Luminosity distance [Mpc] & $d_L$ & $\mathcal{U}_{\text{com. vol.}}(10, 300)$ \\
& & Trigger time [GPS] & $t_c$ & $1240215503.02$ \\
\cline{2-5}
& \multirow{5}{*}{\rotatebox{90}{GW230529}} & Chirp mass [M$_\odot$] & $\mathcal{M}_c$ & $\mathcal{U}_{\rm{comp}}(2.02, 2.04)$ \\
& & Mass ratio & $q$ & $\mathcal{U}_{\rm{comp}}(0.125, 1)$ \\
& & Tidal deformability & $\Lambda_i$ & $\mathcal{U}(0, 5000)$ \\
& & Luminosity distance [Mpc] & $d_L$ & $\mathcal{U}_{\text{com. vol.}}(10, 500)$ \\
& & Trigger time [GPS] & $t_c$ & $1369419318.75$ \\
\hline\hline
\end{tabular}
\end{table}
\renewcommand{\arraystretch}{1.0}

\section{Normalizing flow implementation}\label{sec: app: NF}

In this section, we provide more details on the numerical implementation of the \acp{NF}.
The flows are trained with \textsc{glasflow}~\cite{williams_uofgravityglasflow_2024, nflows}, using the \texttt{CouplingNSF} architecture, which is a type of neural spline flow~\cite{Durkan:2019nsq}.
We use $4$ layers with widths of $256$ neurons and a depth of $3$ blocks for the conditioner networks.
While this results in fairly large networks, we have focused on robustness of the flows across all $18$ configurations shown in Fig.~\ref{fig: neural priors}, and leave further optimization to future work.

We use the procedure outlined in Sec.~\ref{sec: methods: neural priors construction details} to generate ${200 \ 000}$ samples, of which $80\%$ is used as training data, and the remaining $20\%$ as validation data.
We found that training of the flows is improved when trained on the source-frame chirp mass $\mathcal{M}_c^\src$ and mass ratio $q = m_2/m_1$ rather than source-frame component masses.
The training data is rescaled to the domain $[0, 1]$, and we apply a Jacobian factor to the flow's log likelihood calculation during sampling to account for this operation.

The flows are trained with the Adam optimizer~\cite{Kingma:2014vow}, with a fixed learning rate of $10^{-4}$ and a batch size of $1024$ for a maximum of $2000$ epochs, and early stopping is employed if no improvement in the loss is observed after $250$ epochs.

To assess the accuracy of the flows, we evaluate the \acp{JSD}~\cite{Lin:1991zzm} for each parameter between the original distribution of training samples and a distribution generated from the flow.
Across all $18$ physics-informed priors that we consider, we find that the highest \ac{JSD} value is $6.2\times10^{-4}$ bits.
Therefore, we can conclude that the trained \acp{NF} are an accurate representation of the underlying distributions.

The trained flows are then used inside a \textsc{bilby} prior class called \texttt{NFPrior}, which subclasses the \texttt{JointPrior} class.
By design, generating samples and evaluating the density of this \texttt{JointPrior} class are treated by the \ac{NF}.
Certain nested samplers, like \textsc{dynesty}, sample a unit hypercube from which physical samples are obtained by rescaling with the appropriate transformations.
In our case, this is achieved by defining a joint multivariate Gaussian distribution object internally in the \texttt{NFPrior} class of the same dimensionality.
By first mapping samples from the unit hypercube to the unit Gaussian, they can then be transformed into the physical space by using the bijections learned by \ac{NF}.
While this work shows applications with nested samplers, we have verified that the physics-informed priors also work in MCMC-based samplers such as \textsc{jim}~\cite{Wong:2023lgb, Wouters:2024oxj}.

After training, the flows do not necessarily respect the physical boundaries of the training data.
For example, the flow has a negligible, but non-zero, probability to generate samples with a mass ratio above $1$.
We have found that in less than $1\%$ of the \ac{NF} samples, the generated sample is unphysical.
This behavior can be corrected by clipping samples generated from the \ac{NF}, as well as setting the probability of a sample to $0$ if it violates any physical bounds.
While this technically makes the \texttt{NFPrior} class unnormalized (even if the \ac{NF} is, by design), we have computed the normalization constant of one flow with \textsc{dynesty}, finding that the distribution is normalized within the errors of the evidence calculation.
Therefore, we consider the physics-informed priors to be normalized for the purposes of this work.
In the future, this issue can be alleviated by either transforming the flow before fitting, to ensure its support matches the target distribution, or by transforming the data to an unbounded space support before fitting to match the \ac{NF}~\cite{ward_flowjax_constrained}.

\section{Posterior credible intervals}\label{sec: app: parameter values}

Table~\ref{tab: parameter values} shows the median and $90\%$ credible intervals of posteriors of selected parameters between different prior choices.
The uninformed prior refers to the default mass and tidal deformability priors listed in Table~\ref{tab: GW parameters and priors}.
We only present constraints for the source type (\ac{BNS} or \ac{NSBH}) classified for each event, and do not show the constraints for GW190425 using the Gaussian mass distribution, as this hypothesis was decisively ruled out.

{
\renewcommand{\arraystretch}{1.35}
\setlength{\tabcolsep}{4pt}
\begin{table*}[t]
    \centering
    \caption{Comparison of median and $90\%$ credible intervals of posteriors of selected parameters under different prior choices.
    The \ac{NS} mass distributions are uniform (U), Gaussian (G), and double Gaussian (DG).
    The reference model (i.e., with the highest Bayesian evidence) is marked in bold.
    }
    \label{tab: parameter values}
    \begin{tabular}{|l|l|l|c|c|c|c|c|c|c|c|}
\hline
\textbf{Event} & \textbf{Mass} & \textbf{EOS} & $m_1^{\mathrm{src}}$ [$M_\odot$] & $m_2^{\mathrm{src}}$ [$M_\odot$] & $q$ & $\Lambda_1$ & $\Lambda_2$ & $\tilde{\Lambda}$ & $\delta\tilde{\Lambda}$ & $d_L$ [Mpc] \\
\hline\hline
\rule{0pt}{3ex}
\multirow{10}{*}{\rotatebox{90}{GW170817 (BNS)}} & \multicolumn{2}{c|}{Uninformed} & $1.47_{-0.11}^{+0.12}$ & $1.26_{-0.10}^{+0.10}$ & $0.86_{-0.12}^{+0.14}$ & $286_{-286}^{+473}$ & $448_{-448}^{+603}$ & $364_{-234}^{+403}$ & $0_{-170}^{+158}$ & $38_{-11}^{+8}$ \\
\cline{2-11}
 & \multirow{3}{*}{U} & PSRs & $1.51_{-0.06}^{+0.09}$ & $1.23_{-0.07}^{+0.05}$ & $0.81_{-0.09}^{+0.06}$ & $413_{-133}^{+110}$ & $1477_{-334}^{+636}$ & $795_{-71}^{+108}$ & $119_{-43}^{+68}$ & $41_{-8}^{+6}$ \\
 &  & PSRs+$\chi_{\rm{EFT}}$ & $1.52_{-0.07}^{+0.08}$ & $1.23_{-0.06}^{+0.06}$ & $0.81_{-0.07}^{+0.08}$ & $215_{-111}^{+87}$ & $712_{-175}^{+219}$ & $390_{-93}^{+98}$ & $58_{-25}^{+19}$ & $40_{-8}^{+6}$ \\
 &  & PSRs+NICER & $1.57_{-0.07}^{+0.05}$ & $1.19_{-0.03}^{+0.05}$ & $0.75_{-0.05}^{+0.06}$ & $264_{-52}^{+86}$ & $1399_{-342}^{+367}$ & $620_{-82}^{+99}$ & $116_{-38}^{+37}$ & $40_{-8}^{+6}$ \\
\cline{2-11}
 & \multirow{3}{*}{\textbf{G}} & PSRs & $1.41_{-0.03}^{+0.04}$ & $1.32_{-0.03}^{+0.03}$ & $0.93_{-0.05}^{+0.04}$ & $578_{-150}^{+131}$ & $874_{-174}^{+229}$ & $713_{-122}^{+143}$ & $35_{-23}^{+28}$ & $39_{-9}^{+7}$ \\
 &  & \textbf{PSRs+}$\boldsymbol{\chi}_{\mathbf{EFT}}$ & \textbf{$1.42_{-0.04}^{+0.05}$} & \textbf{$1.31_{-0.04}^{+0.03}$} & \textbf{$0.92_{-0.06}^{+0.05}$} & \textbf{$298_{-99}^{+130}$} & \textbf{$495_{-118}^{+173}$} & \textbf{$385_{-85}^{+141}$} & \textbf{$21_{-13}^{+19}$} & \textbf{$40_{-9}^{+6}$} \\
 &  & PSRs+NICER & $1.39_{-0.02}^{+0.02}$ & $1.34_{-0.02}^{+0.02}$ & $0.96_{-0.03}^{+0.03}$ & $641_{-128}^{+186}$ & $826_{-178}^{+277}$ & $729_{-133}^{+219}$ & $21_{-17}^{+22}$ & $38_{-8}^{+7}$ \\
\cline{2-11}
 & \multirow{3}{*}{DG} & PSRs & $1.40_{-0.02}^{+0.03}$ & $1.33_{-0.02}^{+0.02}$ & $0.95_{-0.03}^{+0.02}$ & $687_{-156}^{+140}$ & $960_{-231}^{+183}$ & $814_{-185}^{+143}$ & $29_{-13}^{+21}$ & $40_{-7}^{+6}$ \\
 &  & PSRs+$\chi_{\rm{EFT}}$ & $1.39_{-0.02}^{+0.02}$ & $1.34_{-0.02}^{+0.02}$ & $0.96_{-0.03}^{+0.03}$ & $382_{-111}^{+148}$ & $507_{-152}^{+153}$ & $444_{-131}^{+136}$ & $11_{-9}^{+12}$ & $39_{-7}^{+7}$ \\
 &  & PSRs+NICER & $1.39_{-0.02}^{+0.02}$ & $1.33_{-0.02}^{+0.02}$ & $0.96_{-0.03}^{+0.02}$ & $703_{-153}^{+184}$ & $934_{-200}^{+200}$ & $806_{-184}^{+157}$ & $26_{-17}^{+14}$ & $38_{-8}^{+7}$ \\
\cline{2-11}
\hline\hline
\rule{0pt}{2ex}
\multirow{7}{*}{\rotatebox{90}{GW190425 (BNS)}} & \multicolumn{2}{c|}{Uninformed} & $1.75_{-0.10}^{+0.13}$ & $1.56_{-0.10}^{+0.10}$ & $0.89_{-0.11}^{+0.11}$ & $292_{-292}^{+541}$ & $415_{-415}^{+694}$ & $374_{-333}^{+436}$ & $0_{-173}^{+187}$ & $157_{-65}^{+64}$ \\
\cline{2-11}
 & \multirow{3}{*}{\textbf{U}} & PSRs & $1.81_{-0.09}^{+0.10}$ & $1.50_{-0.08}^{+0.07}$ & $0.83_{-0.09}^{+0.08}$ & $109_{-57}^{+56}$ & $428_{-200}^{+209}$ & $230_{-71}^{+73}$ & $39_{-23}^{+24}$ & $173_{-47}^{+54}$ \\
 &  & PSRs+$\chi_{\rm{EFT}}$ & $1.83_{-0.10}^{+0.12}$ & $1.48_{-0.10}^{+0.07}$ & $0.81_{-0.09}^{+0.09}$ & $58_{-28}^{+40}$ & $259_{-107}^{+174}$ & $132_{-40}^{+44}$ & $23_{-12}^{+21}$ & $187_{-49}^{+48}$ \\
 &  & \textbf{PSRs+NICER} & \textbf{$1.82_{-0.12}^{+0.10}$} & \textbf{$1.48_{-0.07}^{+0.10}$} & \textbf{$0.81_{-0.09}^{+0.11}$} & \textbf{$108_{-56}^{+81}$} & \textbf{$449_{-150}^{+132}$} & \textbf{$232_{-51}^{+57}$} & \textbf{$40_{-23}^{+15}$} & \textbf{$182_{-49}^{+41}$} \\
\cline{2-11}
 & \multirow{3}{*}{DG} & PSRs & $1.93_{-0.08}^{+0.08}$ & $1.41_{-0.05}^{+0.05}$ & $0.73_{-0.05}^{+0.06}$ & $67_{-38}^{+37}$ & $636_{-206}^{+217}$ & $238_{-61}^{+65}$ & $63_{-21}^{+21}$ & $186_{-44}^{+44}$ \\
 &  & PSRs+$\chi_{\rm{EFT}}$ & $1.94_{-0.06}^{+0.07}$ & $1.40_{-0.05}^{+0.04}$ & $0.72_{-0.05}^{+0.04}$ & $51_{-24}^{+23}$ & $408_{-100}^{+125}$ & $156_{-45}^{+37}$ & $37_{-9}^{+10}$ & $190_{-47}^{+41}$ \\
 &  & PSRs+NICER & $1.90_{-0.07}^{+0.08}$ & $1.42_{-0.05}^{+0.06}$ & $0.75_{-0.06}^{+0.06}$ & $96_{-36}^{+41}$ & $690_{-164}^{+188}$ & $283_{-51}^{+45}$ & $66_{-17}^{+18}$ & $186_{-48}^{+41}$ \\
\cline{2-11}
\hline\hline
\rule{0pt}{2ex}
\multirow{10}{*}{\rotatebox{90}{GW230529 (NSBH)}} & \multicolumn{2}{c|}{Uninformed} & $3.65_{-0.85}^{+0.62}$ & $1.43_{-0.24}^{+0.26}$ & $0.39_{-0.13}^{+0.17}$ & -- & $2791_{-1950}^{+2113}$ & $182_{-182}^{+243}$ & $91_{-91}^{+115}$ & $201_{-97}^{+84}$ \\
\cline{2-11}
 & \multirow{3}{*}{U} & PSRs & $3.73_{-0.32}^{+0.28}$ & $1.38_{-0.09}^{+0.10}$ & $0.37_{-0.05}^{+0.06}$ & -- & $698_{-303}^{+344}$ & $41_{-8}^{+8}$ & $21_{-4}^{+4}$ & $238_{-63}^{+61}$ \\
 &  & PSRs+$\chi_{\rm{EFT}}$ & $3.52_{-0.45}^{+0.46}$ & $1.46_{-0.14}^{+0.18}$ & $0.41_{-0.08}^{+0.12}$ & -- & $322_{-202}^{+206}$ & $26_{-7}^{+8}$ & $13_{-4}^{+4}$ & $220_{-64}^{+70}$ \\
 &  & PSRs+NICER & $3.27_{-0.57}^{+0.27}$ & $1.54_{-0.12}^{+0.25}$ & $0.47_{-0.08}^{+0.17}$ & -- & $368_{-291}^{+213}$ & $40_{-14}^{+13}$ & $19_{-7}^{+6}$ & $237_{-67}^{+65}$ \\
\cline{2-11}
 & \multirow{3}{*}{\textbf{G}} & PSRs & $3.86_{-0.26}^{+0.28}$ & $1.34_{-0.08}^{+0.08}$ & $0.35_{-0.05}^{+0.04}$ & -- & $933_{-451}^{+497}$ & $46_{-16}^{+14}$ & $23_{-8}^{+7}$ & $236_{-58}^{+59}$ \\
 &  & PSRs+$\chi_{\rm{EFT}}$ & $3.83_{-0.25}^{+0.25}$ & $1.35_{-0.07}^{+0.07}$ & $0.35_{-0.04}^{+0.04}$ & -- & $582_{-198}^{+261}$ & $30_{-8}^{+7}$ & $15_{-4}^{+3}$ & $242_{-57}^{+59}$ \\
 &  & \textbf{PSRs+NICER} & \textbf{$3.89_{-0.24}^{+0.22}$} & \textbf{$1.33_{-0.06}^{+0.07}$} & \textbf{$0.34_{-0.03}^{+0.04}$} & -- & \textbf{$1000_{-311}^{+321}$} & \textbf{$49_{-9}^{+9}$} & \textbf{$25_{-5}^{+4}$} & \textbf{$235_{-58}^{+59}$} \\
\cline{2-11}
 & \multirow{3}{*}{DG} & PSRs & $3.78_{-0.24}^{+0.23}$ & $1.36_{-0.06}^{+0.07}$ & $0.36_{-0.04}^{+0.04}$ & -- & $890_{-333}^{+341}$ & $49_{-13}^{+12}$ & $25_{-7}^{+6}$ & $237_{-64}^{+63}$ \\
 &  & PSRs+$\chi_{\rm{EFT}}$ & $3.85_{-0.21}^{+0.18}$ & $1.34_{-0.06}^{+0.05}$ & $0.35_{-0.03}^{+0.03}$ & -- & $516_{-205}^{+194}$ & $26_{-8}^{+8}$ & $13_{-4}^{+4}$ & $237_{-71}^{+66}$ \\
 &  & PSRs+NICER & $3.86_{-0.16}^{+0.15}$ & $1.34_{-0.04}^{+0.04}$ & $0.35_{-0.02}^{+0.02}$ & -- & $1026_{-306}^{+254}$ & $52_{-13}^{+11}$ & $26_{-6}^{+6}$ & $229_{-68}^{+65}$ \\
\cline{2-11}
\hline
\end{tabular}
\end{table*}
}

\bibliography{references}{}
\bibliographystyle{apsrev4-1}

\end{document}